\documentclass[12pt,preprint]{aastex}

\usepackage{epsfig}

\shorttitle{IRS SPECTRUM OF PPN}
\begin{document}

\title{A {SPITZER}/IRS SPECTRAL STUDY OF A SAMPLE OF GALACTIC CARBON-RICH PROTO-PLANETARY NEBULAE}

\author{Yong Zhang and Sun Kwok}
 \affil{Department of Physics, University of Hong Kong, Pokfulam Rd., Hong Kong}
 \email{zhangy96@hku.hk; sunkwok@hku.hk}

\and

\author{Bruce J. Hrivnak}
\affil{Department of Physics \& Astronomy, Valparaiso University, Valparaiso, IN 46383, U.S.A.}
 \email{bruce.hrivnak@valpo.edu}
 
\begin{abstract}

Recent infrared spectroscopic observations have shown that proto-planetary nebulae (PPNs) are sites of active synthesis of organic compounds in the late stages of stellar evolution.  This paper presents a study of {\it Spitzer}/IRS spectra for a sample of carbon-rich PPNs, all except one of which show the unidentified 21\,$\mu$m emission feature. 
The strengths of the aromatic infrared band (AIB), 21 $\mu$m, and 30 $\mu$m features are obtained by decomposition of the spectra. 
The observed variations in the strengths and peak wavelengths of the features support the model that the newly synthesized organic compounds gradually change from aliphatic to aromatic characteristics as stars evolve from PPNs to planetary nebulae.

\end{abstract}

\keywords{infrared: stars ---
stars: AGB and post-AGB --- stars: circumstellar matter 
}

\section{Introduction}

Proto-planetary nebulae (PPNs)  represent  a relatively short stage ($\sim10^3$\,yr) when a low- or intermediate-mass star evolves from the asymptotic giant branch (AGB) into the planetary nebula (PN) phase \citep{kwok93, vanwinckel}. PPNs have central stars with B--G spectral type and are not yet hot enough to ionize the surrounding AGB remnant envelope.  As descendants of AGB stars, PPNs  are expected to inherit many of the observed properties of AGB stars, including infrared excess and molecule-line emission from the dust and molecular components of the circumstellar envelopes of AGB stars.  By comparing the spectroscopic properties of the dust component from AGB stars to PPNs to PNs, one sees evidence for a continuous evolution as the result of chemical synthesis and a hardening radiation field \citep{kwok04}.  While the infrared spectra of AGB stars are characterized by silicates or silicon carbide emission features \citep{kwok1997}, the aromatic infrared bands (AIBs) dominate the infrared spectra of carbon-rich PNs.  
How this chemical transition occurs is one of the most interesting question in astrochemistry.  PPNs, being in the evolutionary stage between these two phases, offer important clues to the chemical synthesis of organic compounds in the late stages of stellar evolution.

Specifically, while carbon-rich PNs show strong emission features at 3.3, 6.2, 7.7--7.9, 8.6, 11.3, and 12.7\,$\mu$m due to aromatic compounds, these features are not seen in AGB stars and they only begin to emerge during the PPN phase.  Furthermore, PPNs show aliphatic features at 3.4 and 6.9 $\mu$m which are generally weak or absent in PNs \citep{geb92, kwok99, kwok01}.  Whether this aliphatic-aromatic transition is the result of chemical synthesis or radiative interactions is yet to be determined.  In a comprehensive study of the peak wavelengths of the AIB features, \citet{slo07} found that the peak wavelengths of the 7.7 and 11.3 $\mu$m AIB features are correlated with the temperatures of the central stars, which they interpret to be the result of change in the aliphatic to aromatic content ratio.

The AIB features, although commonly assumed in astronomical literature to be due to polycyclic aromatic hydrocarbon (PAH) molecules, are in fact more likely (at least in the circumstellar environment of late-type stars) to arise from complex, amorphous, impure organic solid-state compounds \citep{kwo09}.  Examples of such amorphous compounds include hydrogenated amorphous carbon \citep[HAC;][]{jones90, scott97}, carbon nanoparticles \citep{hu08}, quenched carbonaceous composites \citep{sak83}, coal or kerogen \citep{pap01}.  Common properties among these candidates are a mixed $sp^2/sp^3$ chemistry, and in the case of coal or kerogen, the presence of elements such as O, S, N, in addition to C and H.  A model of such compounds has been presented by \citet{pen2002}.

To complicate the issues further, it is now known that the infrared spectra of C-rich PPNs contain two unidentified broad emission features at around 21 and 30\,$\mu$m  \citep[see][and references therein]{hri09}. 
The 21\,$\mu$m feature was first discovered in spectra taken by the {\it Infrared Astronomical Satellite} ({\it IRAS}) Low Resolution Spectrometer (LRS) \citep{kwo89}. Based on an {\it Infrared Space Observatory} 
({\it ISO}) study of eight 21\,$\mu$m sources, \citet{vol99} found that they all have the same intrinsic profile and peak wavelength (20.1\,$\mu$m). 
To date, the number of  known 21\,$\mu$m sources is very small, %($\sim25$), 
$\sim$16 in the Galaxy \citep{hri08} and $\sim$8 in the Large and Small Magallenic Clouds \citep{vol10}. 
The carrier of the 21\,$\mu$m feature remains unknown.
A number of candidates, including TiC, AIBs, fullerenes, hydrogenated amorphous carbon (HAC), nanodiamonds etc, have been proposed (see Speck \& Hofmeister (2004) and Posch et al. (2004) and references therein).
The fact that all or almost all the  21\,$\mu$m sources are C-rich PPNs suggests that the carrier of this feature is carbon-based and is associated with a very rapid chemical processing.

The very broad 30\,$\mu$m  feature, usually extending from 25 to 45\,$\mu$m, was first discovered by \citet{for81} in C-rich AGB stars and PNs, and subsequently has been commonly detected in C-rich PPNs \citep{vol02}. An {\it ISO} study of a large sample of evolved stars shows  large variations in the peak wavelength and width of the 30\,$\mu$m  feature among different sources \citep{hon02}.  On the basis of the {\it ISO} spectra of PPNs,
\citet{vol02} found that the feature can be resolved into a broad component at 33\,$\mu$m  and a narrower component at 26\,$\mu$m, although \citet{slo03} argued that the 26\,$\mu$m feature is an artifact caused by an inaccurate calibration.  The origin of the 30\,$\mu$m  feature  is still under debate.  It was first suggested as solid magnesium sulfide \citep[MgS;][]{goe85}.  \citet{hon02} modeled the profile using MgS dust grains with a temperature different from the bulk of dust that contributes to the continuum. They argued that the 26\,$\mu$m feature can be explained by MgS with a different shape distribution.
However, \citet{zha09} found that the available MgS dust mass in the circumstellar envelope is too low to account for the observed feature strength.  Moreover,  the fact that the feature is only detected in C-rich sources and takes up a significant fraction ($\sim20\%$) of the total luminosity of the object seems to favor the more abundant carbonaceous material as possible carrier.

The infrared spectra of C-rich PPNs therefore consist of contributions from several components: a base continuum commonly assumed to be due to thermal emission from amorphous carbon grains, a component due to aromatic and aliphatic compounds, and broad emission features at 21 and 30 $\mu$m.  With the {\it Spitzer} spectroscopic survey of PPNs, we now have a large enough sample to derive the systematic behavior of the spectral features and draw conclusions on their possible evolution during the transition from the AGB to the PN phase.

In this study, we  present {\it Spitzer} spectra of a sample of carbon-rich PPNs, with the goal to investigate the spectral behavior of the AIB, 21\,$\mu$m, and 30\,$\mu$m features. 
In order to derive the intrinsic strengths and profiles of the broad and often overlapping emission features, we need to correctly subtract the continuum under the features.  
A procedure of optimized fitting is used to decompose the observed spectra into individual emission components.
The paper is organized as follows. In Section 2, we briefly describe the data
used in the current study.
The method for spectral decomposition is presented in Section 3.
In Section 4, we report the measurement results and discuss the 
implications of our findings. The last section summarizes the conclusions.

%%xxxxxxxxxxxxxxxxxxxxxxxxxxxxxxxxxxxxxxxx

\section{Data}

The sample includes ten PPNs, all of which except one (IRAS\,01005+7910), are known 21\,$\mu$m sources. 
The observations were carried out with the Infrared Spectrograph (IRS; Houck et al. 2004) on the {\it Spitzer Space Telescope} \citep[{\it Spitzer};][]{wer04} at different epochs between 2004--2008.
Most of the data were obtained from the {\it Spitzer} archive as part of the program No. 20208 (PI: B. Hrivnak) and 93 (PI: D. Cruikshank).   These sources were observed with the Short-High (SH) and Long-High (LH) modules, covering the wavelength range from 9.9--19.6\,$\mu$m  and 18.7--37.2\,$\mu$m, respectively. 
The spectra of IRAS 07134+1005 and  IRAS 20000+3239 (with extremely strong 21 $\mu$m emission) are
not included in our analysis because the 21-$\mu$m feature is saturated in the observations.
We also include the data of three targets (IRAS\,01005+7910, IRAS\,04296+3429, and IRAS\,22223+4327) observed with the short-Low (SL) module in {\it Spitzer}  programs 30036 (PI: G. Fazio) 
and 45 (PI: T. Roellig), which cover a larger wavelength range (5.5--37.2\,$\mu$m). 
These allow the study of the shorter wavelength 6--10\,$\mu$m AIB features. 
In some cases there are discontinuities in the fluxes between the  different modules. This is due to different slit sizes and slight mis-pointing. To ensure that all spectra were on the same flux scale, the regions of spectral overlap are compared and used to scale and combine the  spectra derived with different modules for each object.  In the case of IRAS 19441+2401, the star is outside of the SH aperture so the spectrum would only have recorded part of the extended emission.  Consequently, the short-wavelength part of the spectrum may not be reliable.
Details of the data processing are the same as described in 
\citet{hri09}.  The extracted spectra are displayed in Figure~\ref{fitting}.

\section{Model fitting}

The IRS spectra of all ten sources show strong infrared excess due to dust continuum emission.  Superimposed on the continuum are several strong, broad emission features.  There is no evidence of strong atomic emission lines as one would expect in a PN.  The only narrow features present are molecular lines.  This is consistent with the fact that the central stars of these objects are not hot enough to photoionize the surrounding circumstellar materials.  In this section, we report on our efforts to fit the observed spectra with a combination of dust continuum, dust features, and H$_2$ lines.

For the spectral decomposition, we used the IDL package PAHFIT originally developed by \citet{smi07}  to fit the {\it Spitzer} IRS spectra of nearby galaxies. The model spectra take into account
the contributions from stellar continuum, thermal dust continuum, H$_2$ lines, and AIBs.  We modified this code to include fittings to the 21 and 30\,$\mu$m  features.   The optimal fitting to the observed spectra is achieved through the Levenberg-Marquardt least-squares algorithm.  The fitting results are shown with the spectra in Figure~\ref{fitting}.

In our model, the stellar continuum is assumed to be a 5000\,K blackbody for all the sources. 
While the individual stellar temperatures are known to be mostly in the range of 5250--7250 K,
the actual stellar contribution to the mid-infrared spectra is minimal (see Fig.~\ref{fitting}).
In almost all  cases, the effects of photospheric continuum contributions are negligible.

The dominant contribution to the mid-IR continuum is due to dust thermal emission. A modified blackbody model [$I_\lambda\sim\lambda^{-2}B_\lambda(T)$, where $B_\lambda(T)$ is the blackbody function with a temperature $T$] is adopted to simulate the dust thermal continuum. However, for most of
the sources, we must use two components with different temperatures to get the best fit.
The warm dust component has temperatures from 140 to 180\,K, and the cold component has temperatures from 60 to 80\,K. The only exception is IRAS\,19477+2401, which does not need a warm component (although we note that the short-wavelength spectrum of this object may not be reliable).

Eight H$_2$ rotational lines from S(0)--S(7) are included in the fitting. The line profiles are assumed to be  Gaussian. The H$_2$ 0--0 S(0) para ground-state transition at 28\,$\mu$m  is clearly detected in IRAS\,04296+3429 and IRAS\,22223+4327.
%(Fig.~\ref{hyd}). 
From the spectra, we can see that the H$_2$ lines are much narrower than dust features.  
%For the other sources, the H$_2$ lines are overwhelmed by noise.

Also included in the fitting are all the AIB features in \citet[see their Table~3]{smi07}. Drude profiles [$I_\lambda\sim\frac{\gamma^2}{(\lambda/\lambda_0-\lambda_0/\lambda)^2+\gamma^2}$, where $\lambda_0$ is the central wavelength, and $\gamma$ is the fractional FWHM] are assumed for each feature.
%The adoption of these AIBs and their parameters are based on extensive studies of various objects \citep[see the references in][]{smi07}.
Figure~\ref{fitting} shows that the 7.7, 11.3, and 12.7\,$\mu$m complexes can be well fitted with blended subfeatures. One should bear in mind that these subfeatures do not necessarily represent distinct physical emission bands.

The intrinsic shape of the 21\,$\mu$m feature has been investigated by \citet{vol99}, who found that the profiles are similar among different sources,  all showing ``a sharp rise on the short-wavelength side, beginning at 18.5\,$\mu$m to a peak at 20.1\,$\mu$m, and a gradual decline on the long-wavelength
side to 24\,$\mu$m''. 
\citet{hri09} found the same shape in the {\it Spitzer} spectra they studied and we see similar shapes for the three additional objects (IRAS 01005+7910, 04296+4329, 22223+4327) in our sample
%Their conclusion is supported by the current {\it Spitzer} IRS observations
(see Figure~\ref{fitting}).  To represent this profile asymmetry, we have assumed an asymmetrical Drude profile with the $\gamma$ value in the long-wavelength side three times larger than that in the short-wavelength side.

The very broad 30\,$\mu$m feature is obviously present in the {\it Spitzer} IRS  spectra. \citet{hon02} found that the peak wavelength of this feature shifts from 26\,$\mu$m in some AGB stars to 38\,$\mu$m in PNs. Through the fitting, we find it difficult to reproduce the observations using a single component because the decomposed profile of this feature differ from star to star. 
Although the {\it Spitzer} spectra do not show the two distinct components in the 30\,$\mu$m feature previously deduced from the {\it ISO} spectra, 
%which seem to have been it has been argued that the the presence of two components in the 30\,$\mu$m feature as shown in {\it ISO} spectra is  due to noisy detector bands \citep{hri09}, 
we find it practical to use two Drude profiles  peaking at 26\,$\mu$m and 30\,$\mu$m to represent this feature as they provide better fits to the data.

Dust extinction  is not considered in the present study as its effect is expected to be small in mid-infrared wavelengths.   If present, extinction may reduce the emission flux at 9.7 $\mu$m and 18 $\mu$m due to interstellar silicate absorption.  In some cases, our model cannot reproduce the fluxes around 9.7 $\mu$m and
18 $\mu$m (the model fluxes at these wavelengths are slightly higher than observed),
which might be an indication of the presence of interstellar dust extinction.   Nevertheless, the small discrepancy does not affect our study of dust features.

Although this study is mainly focused on individual features, we explore the possibility
of using principal component analysis (PCA) to study global spectral properties and 
classify PPNs. Our results are presented in Appendix A.

\section{Results and discussion}

We are able to obtain reasonably good fits to the observed spectra using the above model and the results are shown in Figure~\ref{fitting}.
In Table~\ref{dust} we give the integrated fluxes of the continuum and the 21 and 30\,$\mu$m features.
The flux fractions of the  21 and 30\,$\mu$m features to the total IR spectra are also presented in this table.  Table~\ref{AIB} lists the strengths of all the major AIB features.  
Since the SL observations were made only for three objects (IRAS\,01005+7910, IRAS\,04296+3429, and IRAS\,22223+4327), we are only able to obtain the 6.2, 6.9, 7.4, 7.8, 8.3, and 8.6 $\mu$m fluxes for these three objects but not the others.
The feature strengths are derived by integrating the fluxes above the model (star + dust) continuum.  To extract the strengths of dust features, some authors
\citep[e.g.,][]{ber09}  derived the continuum through fitting a spline  function to selected spectral regions free from feature emission. From the PPN spectra shown  in Figure~\ref{fitting}, we can see that the features are  broad and are blended with each other.  Consequently, it is difficult to find a region completely free of dust features to adequately define the baseline.  Given the difficulty in selecting appropriate sections of the spectra to represent the continuum, we believe that  our present method  gives more robust results.

%\subsection{21\,$\mu$m and 30\,$\mu$m  features}

For almost all of the targets, two dust components with different temperatures are required to reproduce the thermal continuum.  The blackbody temperatures used for the fitting are listed in Table \ref{dust}.  The cooler component probably arises from an extended circumstellar envelope left over from the AGB progenitors, and the warm component may originate from a central torus or more recent mass loss.  
Some PPNs have been found to exhibit equatorial tori.   For example, the dust continuum and CO map of the 21 $\mu$m source IRAS\,07134+1005  suggests the presence of torus \citep{kwok2002, meix04,nak09}.  
%High-resolution images at mid- and far-IR wavelengths will be needed to verify this point. 

Tables~\ref{dust} and Table~\ref{AIB} also give the fitting uncertainties of the feature strengths, which are estimated using the full covariance matrix of the least-squares parameters. We can see that the errors are typically less than $20\%$. Adding more blackbody components with temperature of 50--200\,K to the fitting might lead to $5-15\%$ changes in the fluxes of the broad 30 $\mu$m feature, but will have minimal effect on  those of the other features ($< 5\%$).

\subsection{21\,$\mu$m and 30\,$\mu$m  features}

The strengths of the 21\,$\mu$m feature vary among the sources. In the strongest  21\,$\mu$m source (IRAS\,06530$-$0213) in our sample, the fraction of the flux emitted by the 21\,$\mu$m feature to the total emitted infrared flux amounts to $\sim$5$\%$.  Since we fixed the peak wavelength and shape of the  21\,$\mu$m feature in the fitting process, the good agreement between the model and observation supports the suggestion that the intrinsic profile of the 21-$\mu$m feature does not change in different sources \citep{vol99}.
%We also note that the two PPNs with weak cold continuum, IRAS\,19477+2401 and IRAS\,22223+4327,  have weak 21\,$\mu$m feature. 
We also note that the two PPNs (IRAS\,19477+2401 and IRAS\,22223+4327) with weak 21\,$\mu$m features have  no or weak warm dust component. This might indicate that the 21\,$\mu$m feature mostly arises from the warmer torus, although infrared imaging is needed to confirm this suggestion.  
The one object in our sample without the 21 $\mu$m feature is IRAS 01005+7910.  
It has a stellar temperature of 21\,000 K, which is considerably hotter than the other PPNs (see Table \ref{dust}).
The presence of the 21 $\mu$m feature in PPNs but not in AGB stars or PNs has previously been interpreted as an indication of the evolution of the feature due to chemical synthesis or changing excitation conditions.

The 30\,$\mu$m feature is present in all of the PPNs. Its flux typically accounts for $\sim$$20\%$ of the total infrared flux in the PPNs (they range from 8  to 28$\%$).  From Figure~\ref{fitting} one can see that the 30 $\mu$m feature has an asymmetric profile with a sharp  blue wing and a broad red tail. 
Although the presence of two separate components for this feature as seen in the {\it ISO} spectra has been shown to be unreal \citep{hri09}, we find it helpful to use two components to fit this broad feature because a single component of fixed shape and central wavelength cannot give a satisfactory fitting. We should note that the two components do not necessarily mean two individual features. We find that the relative flux ratios of the two components vary among sources.  
This representation of the changing shape and peak is consistent with the findings of
\citet{hon02} who simulated the MgS emission feature and found that the peak position changes with the grain shape and temperature.

\citet{ber09} investigated the {\it Spitzer}/IRS spectra of 25 PNs in the Magellanic Clouds (MCs) and found that about half of them possess the 30 $\mu$m feature. They did not find a clear relation between the presence of this feature and nebular sulfur abundance, which is used to argue against the suggestion of MgS as possible carrier. 
The strengths of the 30 $\mu$m features in MC PNs \citep[see Figure~6 in][]{ber09} are similar to those in MW PNs, suggesting that the different metallicity of the two galactic systems is not a major factor in the existence of the 30 $\mu$m feature.  However, when comparison is made to the strengths of the feature in MW PPNs as presented in this study, one sees that the feature is stronger in PPNs.  This suggests that the 30 $\mu$m carrier is gradually being destroyed or not excited in the PN phase.
If MgS is the carrier, we expect the amount of sulfur released into the gaseous nebula will increase as the star evolve to a PN. However, the low sulfur abundance found in PNs  %[What PNs?  MW or MC; C-rich or O-rich?] 
\citep[e.g.][]{pot06,ber08} does not seem to  support this idea (although there exist other explanations for the `sulfur anomaly'; see Henry et al. 2006). 
The decline of the strength of the 30 $\mu$m feature is more likely to be the result of changing physical or chemical conditions.

\subsection{AIBs}

%%%%%Did you try to quantify the lack of correlation between the strength of 21 um and AIBs?  Graph these some how. Or how about adding a column to Table 2 of 11.3/21 intensity ratio.  You do state 2 cases where they are each low and make a suggestion based on this. 

The AIB features are prominently present in all of the PPNs in this sample and the peak wavelengths of the features detected are given in Table~\ref{wave}.
The profiles and central wavelengths of AIBs are known to vary in different astronomical sources \citep[see, e.g.][]{peeters02}.  %There are several factors affecting  the position of AIB peak wavelengths, including molecular size, molecular symmetry, and molecular heterogeneity \citep{peeters02}.  The variations of peak positions of AIB features in different objects reflect different physical conditions.
Almost all of the sources have the 11.3 $\mu$m feature peaking at the same wavelength, 11.39$\pm$ 0.03 $\mu$m.  The exceptions are IRAS 22574+6606 and IRAS 01005+7910, the latter of which is a much hotter source and does not possess the 21 $\mu$m feature.

In order to examine the features in detail, we have plotted in  Figure~\ref{spe} the continuum-subtracted spectra of the three PPNs observed in the IRS mode SL 6--10 $\mu$m region together with the spectra of the young PN IRAS 21282+5050.  The peak wavelengths of the aromatic and aliphatic emission features are marked.  
The 6.2\,$\mu$m aromatic feature is detected in all three PPNs that were observed in this wavelength region. 
The 6.9 and 7.25 $\mu$m aliphatic features are particularly distinct in IRAS 04296+3429, whereas the 8.6 $\mu$m aromatic feature is relatively weak in that object.
In IRAS 22223+4327, the 6.9 $\mu$m is not obvious from looking at the spectrum alone, but from the fitting results, the integrated strength of this feature is quite high due to its width.

One can compare the 6.2\,$\mu$m/11.3\,$\mu$m intensity ratio to that found in PNs in the MW, LMC, and SMC 
\citep[see their Fig.~2]{ber09}. 
For IRAS 04296+3429 and IRAS 22223+4327, the ratio is low, 0.45 and 0.47, respectively, while for the hotter PPN IRAS 01005+7910, the ratio of 1.40 is similar to that found for most of the PNs.  
An examination of the shorter wavelength features also shows shifts in the peaks of the 
features between IRAS 01005+7910 and the other two sources observed and also the appearance of a small shift in the peak of the broad 7.7-8.2 $\mu$m complex between IRAS 22223+4327 and 04296+3429. 
Also seen is a difference in the peak at 7.60 $\mu$m for the hot PPN IRAS 01005+7910 and 7.85 $\mu$m for the PN IRAS 21282+5050.  There seems to be a general trend for the peak wavelength of the 7.7 $\mu$m feature to shift to shorter wavelengths with evolution.

In Figure~\ref{peak}, we have plotted the peak wavelengths of the 11.3\,$\mu$m feature as functions of the fractional strengths of the  of the 21\,$\mu$m feature.  The only two objects with redshifted 11.3 $\mu$m are the ones with weak or no 21 $\mu$m feature.
%It shows that almost all of the objects have the 11.3 $\mu$m feature peaking at approximately the same wavelength, $11.39\pm0.03$ $\mu$m.  The exceptions are IRAS 22574+5435 and and IRAS 01005+7910.  As noted earlier, IRAS 01005+7910 is a much hotter sources and does not possess the 21 $\mu$m feature.
The peak wavelengths of the 11.3 $\mu$m feature as a function of central star 
temperatures are plotted  in Figure~\ref{peak2}.  For comparison, we have also included in Figure~\ref{peak2} the values for Ae/Be and other C-rich stars from \citet{slo07} and  \citet{keller08}.  Although these objects do not belong to a homogeneous sample, this figure seems to suggest that  the peak of the 11.3\,$\mu$m feature occurs at shorter wavelengths at higher central-star temperatures.  
%These two plots are consistent with the suggestion that as the star evolves to higher temperatures, the peak of the 11.3 $\mu$m feature shifts to shorter wavelengths and the strengths of the 21 $\mu$ feature weakens.
It has been suggested that the sources with most redshifted AIB features are exposed to weaker UV radiation fields and are less processed \citep{slo07}.  While we note that the amount of UV radiation background in our PPN sample is minimal,  our results are consistent with the trends as reported by \citet{slo07}.
Proposed explanations for the wavelength change of the AIB features include nitrogen-inclusion, variations in the fraction of carbon isotope, size distribution of the grains, ionization state, and the content ratio of aliphatic to aromatic bands \citep[see, e.g.,][and references therein]{ber09}.  We suggest that this shift of the peaks of the AIB features to shorter wavelengths is related to the relative strengths of the aliphatic to aromatic components as well as growth of the sizes of the aromatic rings as the star evolve from PPN to PN \citep{kwok01}.
Photochemistry may be the underlying cause of these changes.

%%%%%(10) 4.2, par4: Fig 4 (or 3): I don't think that this makes a case for peak shift compared with flux – 01005 is much hotter and I think that the stronger case, supported by other observations of hot objects, is that 21 um feature gets very weak or goes away by the time one gets to early B.  Similarly the peak wavelength shifts at hotter temperatures.  Then the question is reduced to whether the shift in 22574 is telling us something – Temperature?

The 11.3 $\mu$m AIB feature is identified with the C$-$H out-of-plane bending mode of aromatic compounds.  However, when the number of rings in the aromatic units is small, there can be 2 (duo), 3 (trio), or 4 (quarto) adjacent CH groups  and their out-of-plane bending mode frequencies can be slightly different \citep{hudgins99}.  Generally, the duo-CH bending mode is in the 11.6-12.5 $\mu$m range, the trio-CH mode is in the 12.4-13.3 $\mu$m, and the quarto-CH mode in the 13-13.6 $\mu$m range.  While the 11.3 $\mu$m solo-CH mode is dominant in PNs, the other modes are seen to be relatively stronger in PPNs \citep{kwok99}.  This observation has led to the suggestion that the aromatic units grow in size as the star evolve from PPNs to PNs \citep{kwok04}.  With our present fitting exercise, it is possible to quantify this change.  In Figure~\ref{duo} we plotted the ratios of the sum of the strengths of the duo, tri, and quatro modes to the solo mode as a function of central star temperature.  It can be seen that the strengths of the duo, trio and quatro modes decrease relative to the solo mode as the star evolves to higher temperatures, supporting the scenario that the aromatic units get larger with evolution.

The derived fluxes from each of the AIB features as well as their total contribution to the observed infrared fluxes are given in Table \ref{AIB}.  Comparing the last column of Table \ref{AIB} to the values in columns 8, 10, 12 in Table \ref{dust}, we can see that the contribution of the AIB features to the total emitted infrared flux is much smaller than that of the 30\,$\mu$m feature and is of the same order (a few \%) as  that of the 21\,$\mu$m feature.  In some cases, the flux of the 11.3 $\mu$m feature alone is larger than (e.g., in IRAS 05341+0852 and IRAS 07430+1115), or about equal to (as in IRAS 22574+5435)  that of the 21 $\mu$m feature .  
%An inspection of Figure~\ref{fitting} suggests that there is no obvious correlation between the strengths of the 21\,$\mu$m feature and those of the AIBs. We note, however, that the two PPNs with weak AIB emission (IRAS 19477+2401 and IRAS 22223+4327) also have weak 21\,$\mu$m emission.  This may suggest that the AIBs and the 21\,$\mu$m carriers share a common relationship in chemical synthesis or in mechanism of destruction.

\citet{cer09} found that about 40$\%$ of their post-AGB star sample exhibit both AIBs and silicate features, and suggested that the spatial segregation of different dust populations in the envelopes can lead to the mixed chemistry and a disk/torus can serve as a reservoir preserving O-bearing molecules. However, we do not detect silicates features in our (C-rich) sources, suggesting that the 21\,$\mu$m sources do not favor the formation/survival of O-bearing compounds.

\subsection{15-20 $\mu$m plateau feature}

A 15--20\,$\mu$m plateau emission feature is commonly seen in PNs and other objects with a hot central star and has been attributed to C-C-C bending modes of aromatic rings \citep{van00}. 
For example, this feature is seen in IRAS 06556+1623 and IRAS 18442$-$1144 and several other post-AGB stars \citep{cer09}.
%\citep[see ]{van00}.  
In most of our sources, this region is difficult to examine because of overlap with the blue wing of the 
21 \,$\mu$m feature.  However, in  IRAS\,01005+7910 this is seen as a prominent feature, as shown in Figure~\ref{plateau}.
Integrating the fluxes between 15.2 and 20\,$\mu$m, we obtain an integrated strength of 8.5$\times10^{-13}$ W/m$^2$, leading to strength
ratios relative to the C-C stretching mode at 6.2\,$\mu$m of 22 ($I_{plateau}/I_{6.2}$) and the C-H out-of-plane bending model at 11.3\,$\mu$m of 31 ($I_{plateau}/I_{11.3}$). The plateau/6.2,11.3 strength ratios are much larger than the values ($\le$5) found in the sample of \ion{H}{2} regions, Ae/Be stars, and PNs studied by \citet{van00}.  
Whether this is an evolutionary trend or particular to IRAS 01005+7910 is difficult to say based on only one case.
In this respect, it would be good if one could obtain 15--20\,$\mu$m spectra of other C-rich PPNs without a strong 21 $\mu$m feature.
%While this is a small sample of one object, it suggests that PPNs may provide a favorite environment for the 15--20\,$\mu$m plateau emission.  

%A prominent  15--20\,$\mu$m plateau emission feature is detected in the non-21\,$\mu$m source, IRAS\,01005+7910, as shown in Fig.~\ref{plateau}.
%This emission plateau has been commonly seen in young stellar objects, compact \ion{H}{2} regions, and PNs  and has been attributed to C-C-C bending modes of aromatic rings \citep{van00}. Integrating the fluxes between
%15.2 and 20\,$\mu$m, we obtain an integrated strength of 8.5$\times10^{-13}$ W/m$^2$, leading to strength
%ratios relative to the C-C stretching mode at 6.2\,$\mu$m ($I_{plateau}/I_{6.2}=$22.1) and the C-H out-of-plane bending model at 11.3\,$\mu$m ($I_{plateau}/I_{11.3}=30.9$). The plateau/6.2,11.3 strength ratios are much larger than those found in the sample of \citet{van00}, suggesting that PPNs may provide a favorite
%a favorite environment for the 15--20\,$\mu$m plateau emission.  Whether the plateau is present in 21\,$\mu$m sources is unclear as it falls within the blue wing of the 21\,$\mu$m feature.

\subsection{15.8 $\mu$m feature}

In a previous paper \citep{hri09}, we called attention to a moderately broad emission feature at 15.8 $\mu$m seen in the {\it Spitzer} spectra of several carbon-rich PPNs with the 21 $\mu$m feature.  
This feature is seen in the several additional spectra included in this study.
While a weak feature at 15.9\,$\mu$m has been detected in the reflection nebula
NGC\,7023 \citep{wer04} and some star-forming galaxies \citep[see Fig.~5 of][]{smi07},
the feature seen in the PPNs in the present study is significantly broader.  We noted there was a suggestion of a correlation with the strength of the 21 $\mu$m feature in the sense that the two sources with the strongest 21 $\mu$m features also had the strongest 15.8 $\mu$m features (IRAS 06530$-$0230 and 23304+6147).  We can examine this more quantitatively in this study with the measurement of the strength of this feature in these sources (see Table 2) and the inclusion of two other 21 $\mu$m sources (IRAS 04296+4329 and IRAS 22223+4327).

Fig.~\ref{vs1521} displays the 15.8\,$\mu$m fluxes versus the 21\,$\mu$m fluxes for the PPN sample. With the exception of IRAS\,04296+3429, there seems to be a positive correlation between the two fluxes, although
more data, specially those of strong 21\,$\mu$m sources, are required to confirm this conjecture. Both 15.8 and 21\,$\mu$m features have an asymmetrical profile. In Fig.~\ref{cor1521}, we compare the profiles of the two features.
While in all the cases the 15.8\,$\mu$m feature is narrower than the 21\,$\mu$m feature, they are similar in shape: both have a steeper blue rise and a shallower red tail. 
%This can be explained by the anharmonicity of vibrational modes which results in slight red shift of the emission from higher vibrational states \citep{barker87}. 

\subsection{Molecular lines}

In addition to the above broad features, there are also molecular lines in the spectra. 
 \citet{hri09} have detected the 13.7 $\mu$m $\nu_5$ line of C$_2$H$_2$ in five PPNs, one (IRAS 22574+6609) in absorption, three (IRAS 05341+0852, IRAS 06530$-$0213, and IRAS 07430+1115) in emission, and one (IRAS 23304+6147) in both absorption and emission (P Cygni profile).
From our examination of the spectra, the C$_2$H$_2$ could also be present in emission in IRAS 04296+3429.
C$_2$H$_2$ is the most abundant molecular species after H$_2$ and CO in carbon-rich envelopes. In our sample, about 70$\%$ PPNs exhibit the C$_2$H$_2$ feature.  IRAS 05341+0652 has the strongest C$_2$H$_2$ emission. We do not find correlation between the fluxes of C$_2$H$_2$ and those of other features.

Furthermore, we can unambiguously identify the H$_2$ 0--0 S(0) line at 28 $\mu$m in IRAS\,04296+3429 and IRAS\,22223+4327 and their spectra are shown in Figure.~\ref{hyd}.
However, the expected next strongest H$_2$ line ortho ground-state  $v=0-0$ S(1) line at 17 $\mu$m is not detected.  We do not find any correlation between the strengths of dust features
and those of molecular hydrogen emission. IRAS\,04296+3429 shows relatively strong AIB and 21\,$\mu$m features, but these features are weak in the spectrum of IRAS\,22223+4327.  Table~\ref{h2} gives the fluxes of the S(0) transition and the flux upper limits of the S(1) transition estimated from
the 3$\sigma$ flux noise level. These values allow us to determine the upper 
limits of excitation temperatures in the two objects. Under an assumption of 
local thermodynamic equilibrium (LTE),
we have

\begin{equation}\label{hh}
\frac{N_{v,J}}{g_J} = N_{\rm tot}\frac{\exp(-E_{v,J}/kT_{\rm ex})}{\Sigma_{v',J'}
g_{J'}\exp(-E_{v',J'}/kT_{\rm ex})},
\end{equation}
where $N_{v,J}$ is the population of the upper level ($v,J$), $g_J$ is
the level degeneracy under the assumption of ortho/para ratio = 3,
($g_J=(2J+1)$ for even $J$ and $3(2J+1)$ for odd $J$),
$N_{\rm tot}$ is the total H$_2$ density, $E_{v,J}$ is the
energy of the upper level, $T_{\rm ex}$ is the excitation temperature,
and $k$ is Boltzmann's constant. The flux of a given transition,
$F$, is proportional to $N_{v,J}$ ($F\propto N_{v,J}A/\lambda$, where
$A$ is the radiative transition probability). From Equation~(\ref{hh}) we have

\begin{equation}
T_{\rm ex}=\frac{E_{v,J_1}/k-E_{v,J_0}/k}{\ln(F_0\lambda_0 g_{J_1}A_1)-\ln(F_1\lambda_1g_{J_0}A_0)},
\end{equation}
where the subscripts, 0 and 1, represent the transitions S(0) and S(1), respectively.
Taking the $E_{v,J}$ and $A$ values from \citet{ros00}, we derive
the $T_{\rm ex}$ upper limits of  56\,K and 51\,K for
IRAS\,04296+3429 and IRAS\,22223+4327, respectively. 
The extremely low excitation temperatures of H$_2$ seem to suggest that
the contributions from UV pumping and shock waves
to the excitation of H$_2$ are insignificant in the two PPNs.

This is consistent with previous observations of the H$_2$ $v=1-0$ S(1) line at 2.12 $\mu$m of these two and other PPNs \citep{kel05}.
It was found that H$_2$ emission was seen in F-G spectral type PPNs only if they were clearly bipolar in morphology with a relatively dense equatorial dust torus and a bipolar outflow, and that the H$_2$ emission in these cases was collisionally excited. 

\section{Concluding remarks}

We have investigated the mid-IR spectral features -- AIBs, 15-20\,$\mu$m plateau, 15.8\,$\mu$m, 21\,$\mu$m, and 30\,$\mu$m -- in a sample of PPNs using their {\it Spitzer}/IRS spectra. The strengths of these emission features are determined using a decomposition method.  We find that the 21 and 30 $\mu$m features are strong in these PPNs, contributing up to 5 and 30\%, respectively, of the total infrared fluxes observed.
A prominent 15-20 $\mu$m broad emission plateaus is detected in IRAS 01005+7910.  The 15.8 $\mu$m emission feature, seen in almost all our objects, is found to have similar intrinsic emission profile as the 21 $\mu$m feature.  
%[TO ADD - Statement regarding feature strengths and peak wavelengths.]  
Molecular hydrogen lines are observed in two of the PPNs, showing the presence of the neutral circumstellar component.  Most of the PPNs require two dust components to fit their underlying continua.  We suggest that the cold component is due to an extended dust halo probably left over from the AGB phase, and a warmer component is due to a dust torus in the central region of the PPN.  This suggestion can be tested by future infrared imaging observations.

%The changes of peak wavelengths of AIB features are confirmed to be related to stellar evolution. With increasing UV radiation field, the 21\,$\mu$m features and  aliphatic bonds are destroyed, and the AIB features are shifted to shorter wavelengths.

The fact that the duo, trio, and quarto C-H out-of-plane bending modes decline in strength with evolution relative to the solo mode is consistent with the suggestion that the aromatic component grows in size as the star evolves to higher temperatures.  There is also evidence for the gradual shift of the peak of the 11.3 $\mu$m feature to the blue with evolution.  The 21 and 30 $\mu$m features, which carriers are unknown, are shown to decline in strength with evolution.  These observations will help us to constrain future chemical evolution of organic synthesis in the late stages of stellar evolution.

\acknowledgments

This work was supported by a grant to SK from the Research Grants Council of the Hong Kong Special Administrative Region, China (Project No. HKU 7020/08P)
and a grant to YZ from the Seed Funding Programme for Basic Research in HKU (200909159007).
%BJH acknowledges funding from .....

\appendix

\section{Principal component analysis}

We also investigated the variety of the PPN spectra using the method of principal component analysis (PCA). PCA is a mathematical tool which has been commonly applied to spectral classification  \citep[e.g.][and references therein]{fra92,suzuki06}.  The basic idea of PCA is to obtain new independent variables, called the principal components (PC), through orthogonal transformation.  The new variables are a linear combination of the original data.  The relative contributions to the spectrum variance of each PC can be determined. The PCA technique can reduce the multidimensionality of the data and enables one to study the correlations between different spectral properties \citep[see,][for a more detailed description of PCA]{fra92}.  This analysis has the potential to give new insight into the spectral features.

In our work, the common wavelength coverage (10--35\,$\mu$m) of the ten PPN spectra was considered for PCA. The formulation of PCA can be found in many papers \citep[e.g.][]{ronen99}. We first subtracted the continuum and normalized each spectrum by its integrated flux. The mean spectrum is obtained by

\begin{equation}
<F(\lambda_j)>=\frac{1}{N}\sum_{i=1}^{N} F_i(\lambda_j),
\end{equation}
where $N=10$ is the number of the PPN spectra, and $j=1, ..., M$
is the number of the wavelength channels. The data set can be
denoted by a $N\times M$ matrix

\begin{equation}
X_{ij}=F_i(\lambda_j)-<F(\lambda_j)>.
\end{equation}
The eigenvector and eigenvalue of its covariance matrix, respectively, give 
the new axis (PCs) and the variance along the new axis. The PCs are
ordered according to their relative contributions to the spectrum variance,
such that the first PC accounts for the maximum variance.

Figure~\ref{pca} shows the mean spectrum and the first three PCs (PC1, PC2,
and PC3). The first three PCs account for 79\% of the variance in the spectra. The mean spectrum mainly exhibits three features: the AIBs around 10--15\,$\mu$m, the 21\,$\mu$m feature,  and the 30\,$\mu$m
feature. PC1 accounts for 37\% of the variance. Compared to other PCs, the most prominent feature of PC1 is the dip 
around 18\,$\mu$m. Thus, it can modulates the strengths of the 21\,$\mu$m feature and the emission from 12--15\,$\mu$m. PC2 accounts for 29\% of the variance. It displays the modulation of the AIB features from 10--15\,$\mu$m, and especially the 11.3 $\mu$m feature.   Since PC2 can also modulate the flux of 12--15\,$\mu$m emission, PC1 particularly displays the modulation in  the flux of the 21\,$\mu$m feature. PC3 account for 13 percent of the variance. It reflects the variation in the relative fluxes of individual AIB features. We can conclude
that the variance in this spectrum sample is mainly caused by the 21\,$\mu$m and AIB features, but not by the 30\,$\mu$m feature.

Figure~\ref{pca_wei} shows the projections of the PPN spectra on to the first
two PCs (denoted by pc1 and pc2). The values of pc1 and pc2 represent the
contribution of each component to the difference between individual PPN 
spectrum and the mean spectrum. We can find that the non-21\,$\mu$m
source, IRAS\,01005+7910, has the lowest pc1 value, 
and the strongest 21 $\mu$m source (IRAS\,06530-0213, IRAS\,23304+6147, IRAS\,04296+3429) have the largest pc1 values. If the 21 $\mu$m  feature arises from the warm torus(see Section~4), the value of pc1 could represent the content of warm dust in the torus. The value of pc2 reflects the strengths of AIBs with respect to the mean spectrum. The sources
with weak AIB features (IRAS\,22223+4327 and IRAS\,19477+2401)  
have the largest pc2 values while the sources with strong AIB features (IRAS\,05341-0852, IRAS\,22574+6609) have the lowest pc2 values. Therefore, PCA allows us to objectively
and quantitatively classify the spectra of PPNs. However, the current
sample is too small for the analysis of higher-order PCs. With a
larger PPN sample, it would be intriguing to use PCA to study the global 
spectral behavior and the relation between the fluxes of different AIB 
features.

\clearpage

\begin{figure*}
\begin{center}
\begin{tabular}{cc}
\resizebox{85mm}{!}{\includegraphics{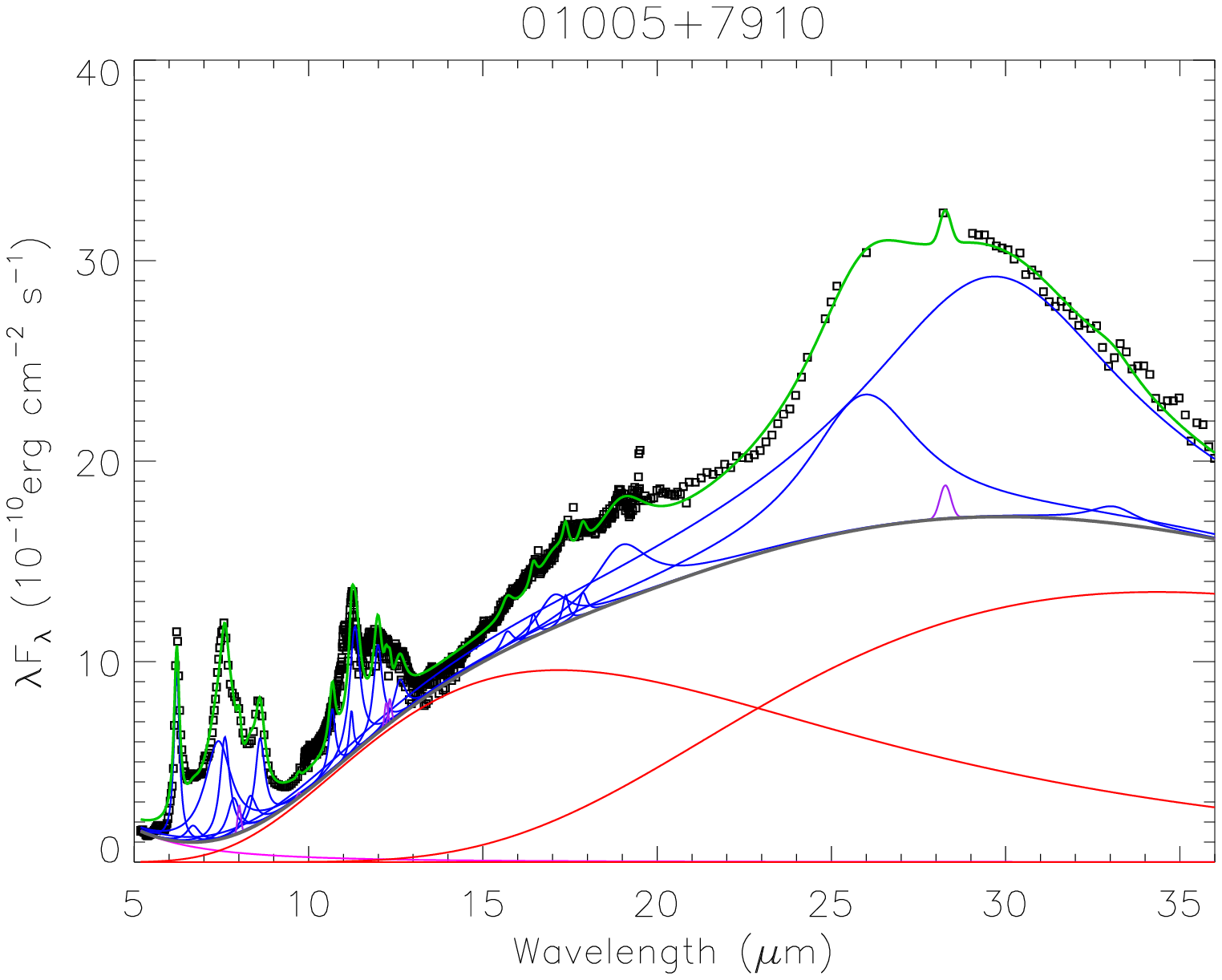}} & \resizebox{85mm}{!}{\includegraphics{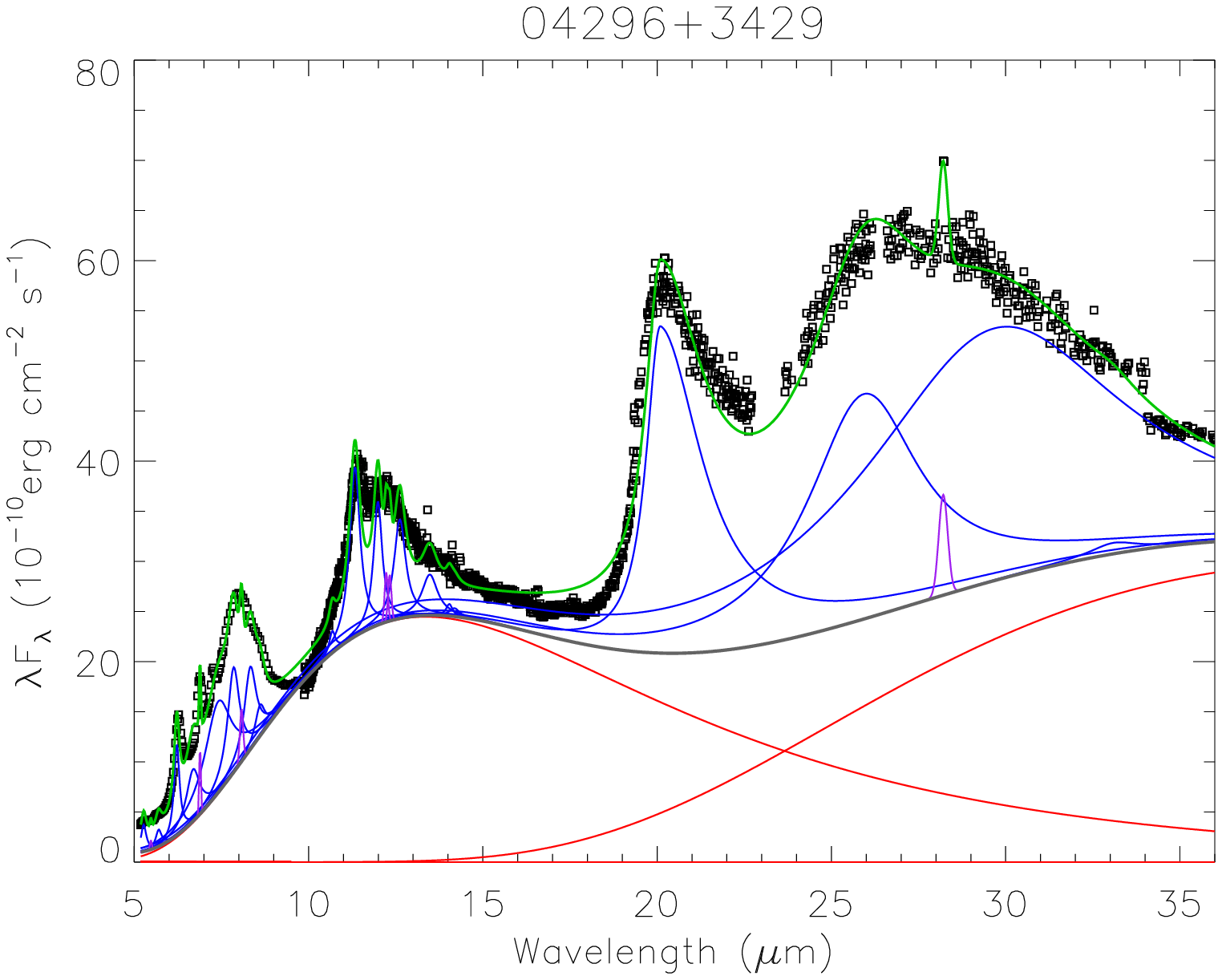}} \\
\resizebox{85mm}{!}{\includegraphics{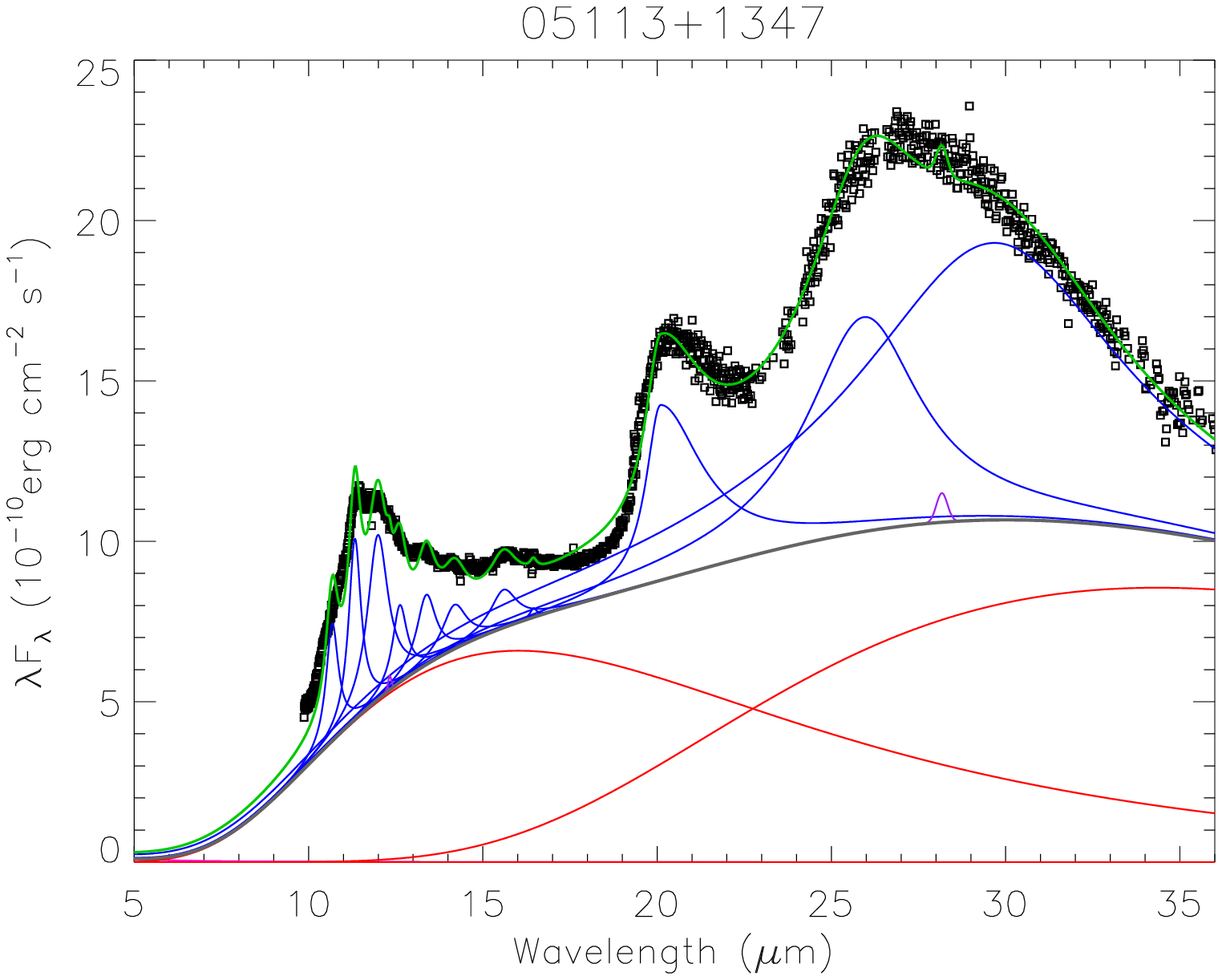}} & \resizebox{85mm}{!}{\includegraphics{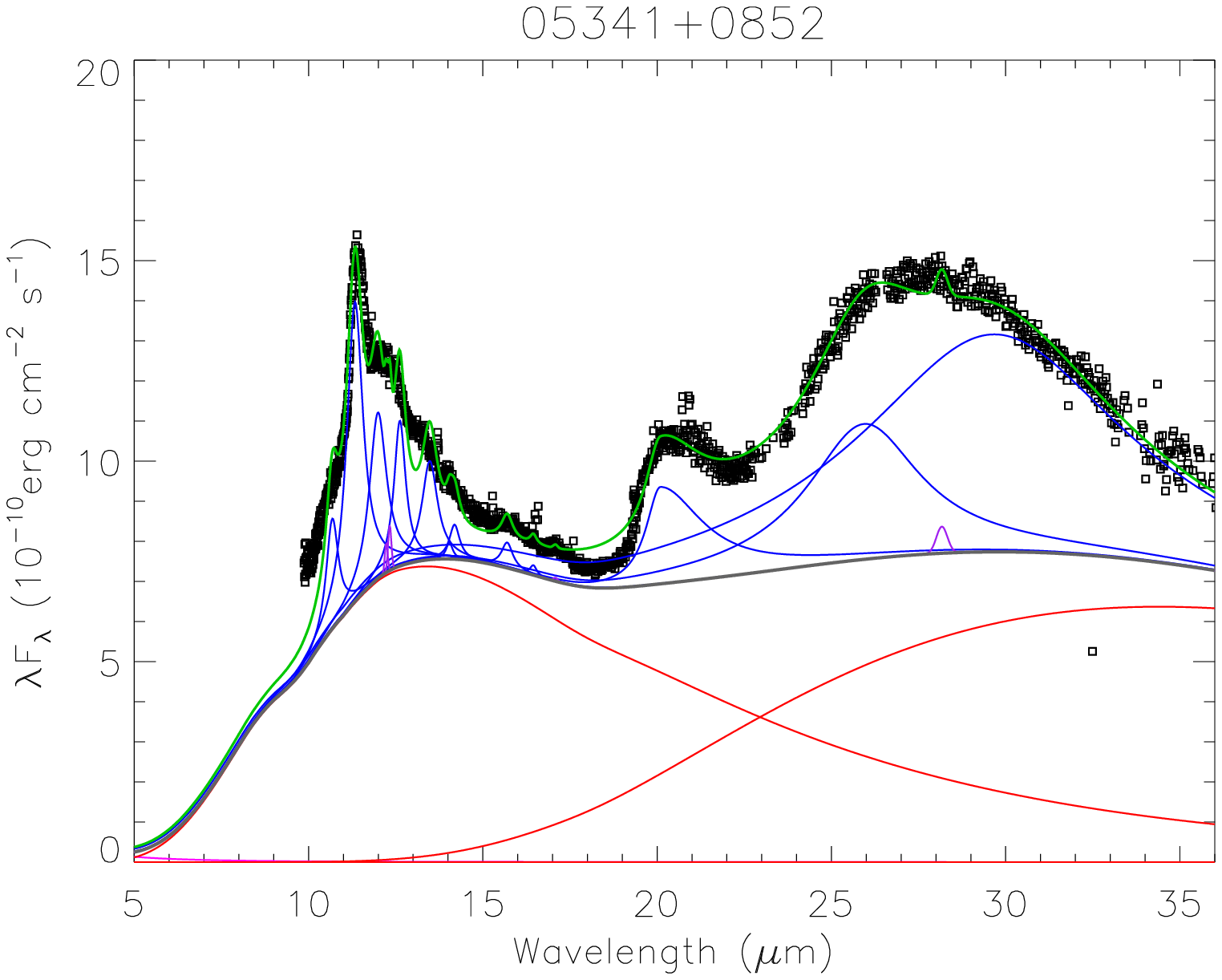}} \\
\resizebox{85mm}{!}{\includegraphics{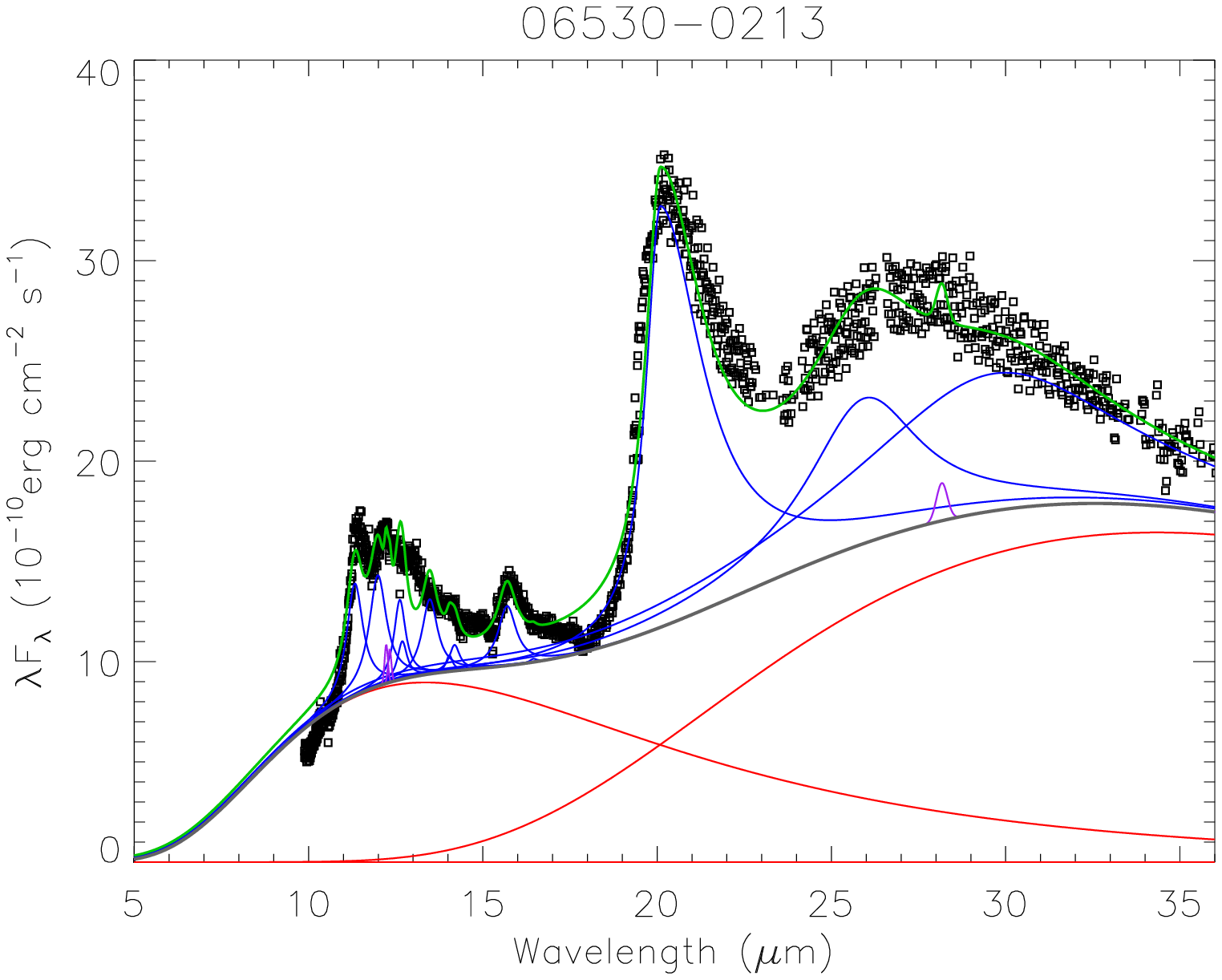}} & \resizebox{85mm}{!}{\includegraphics{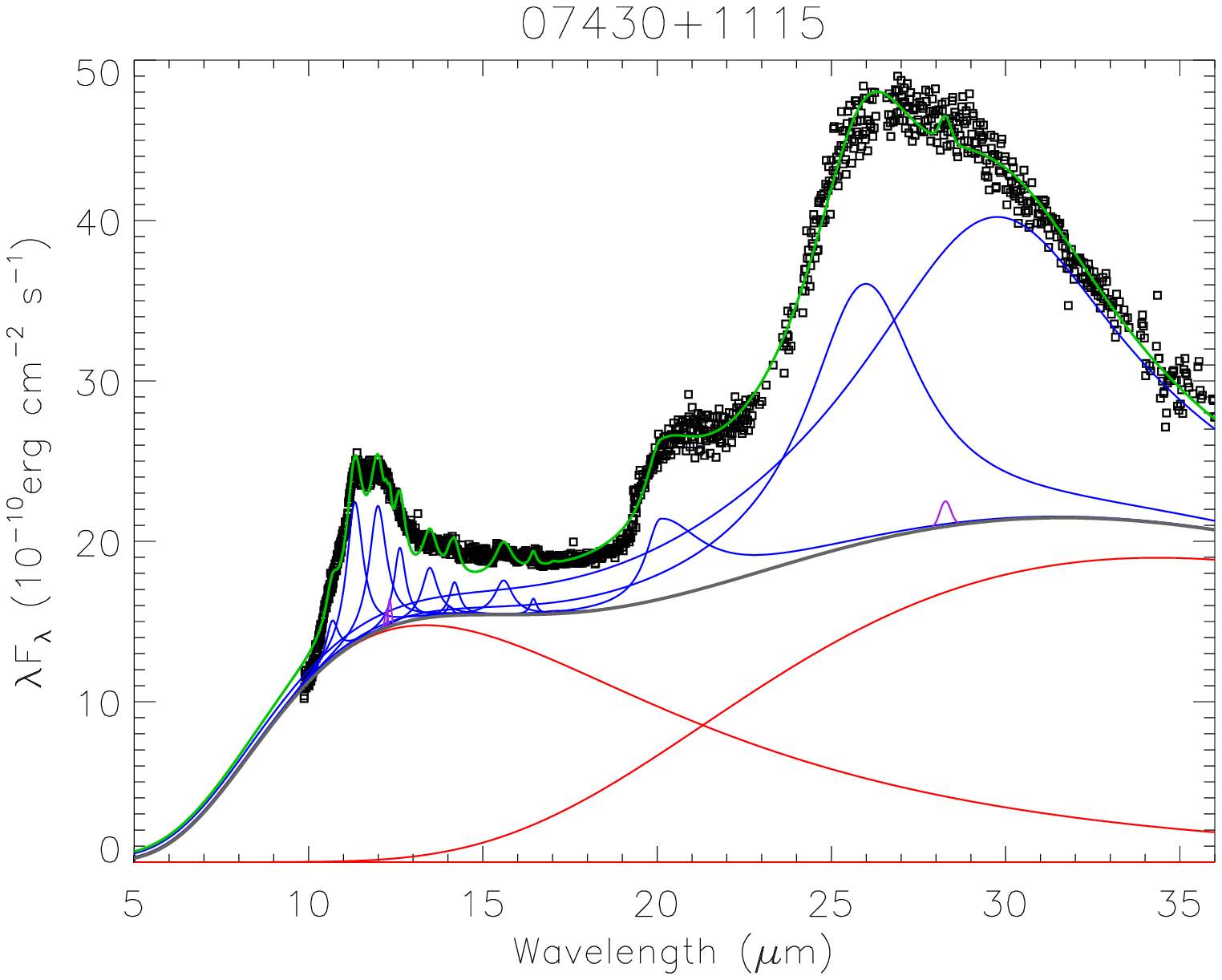}} \\
\end{tabular}
\end{center}
\caption{Decomposition of the IRS spectra. The black squares are the observations. The red solid lines show the thermal dust components. The stellar continuum, although included in the plots, are too weak to be seen.
The gray lines show the total of the stellar and dust continuum components.  The dust features and H$_2$ lines are indicated by the blue and violet lines, respectively. The green solid lines are total model spectra.
\label{fitting}
}
\end{figure*}

\addtocounter{figure}{-1}
\begin{figure*}
\begin{center}
\begin{tabular}{cc}
\resizebox{85mm}{!}{\includegraphics{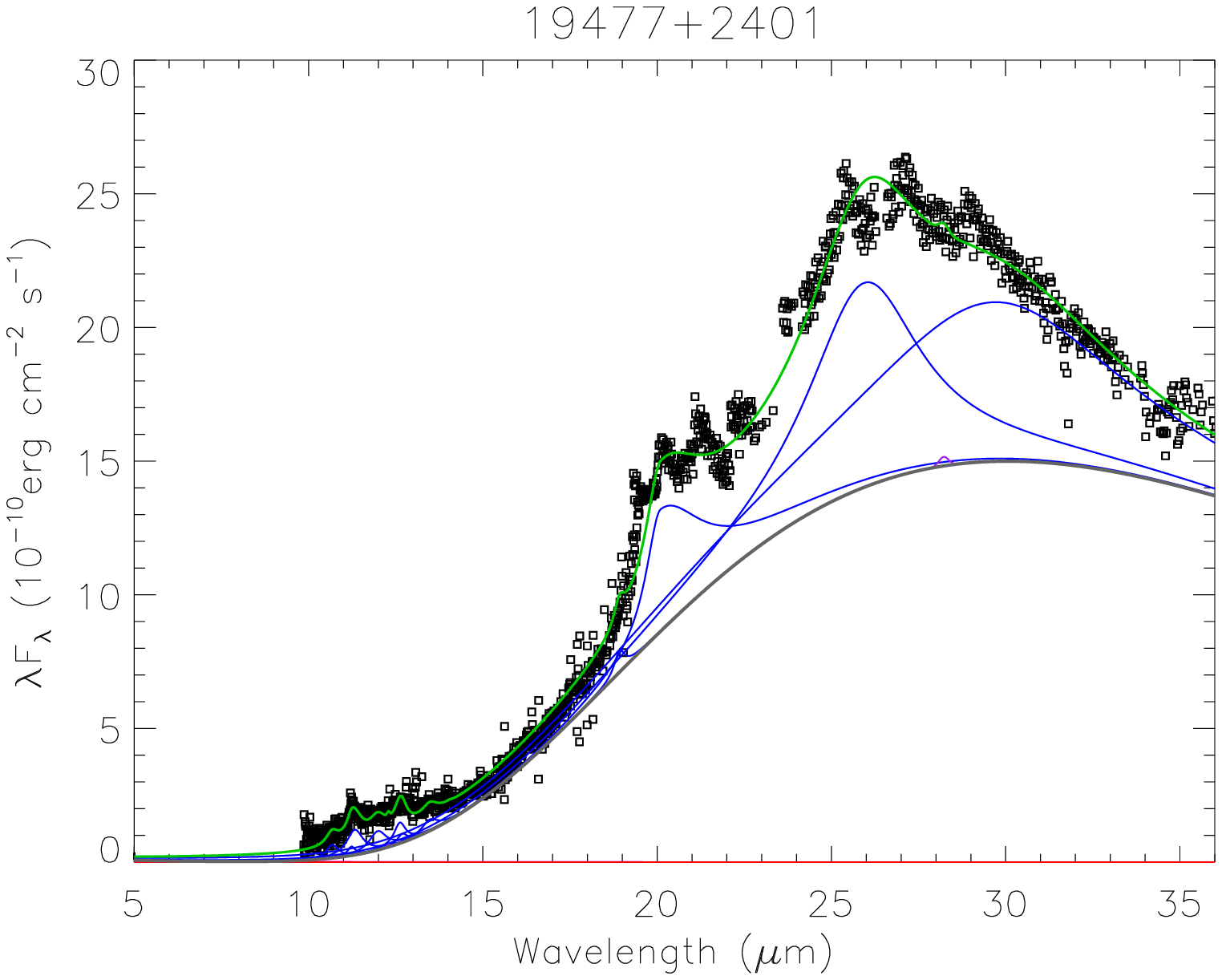}} & \resizebox{85mm}{!}{\includegraphics{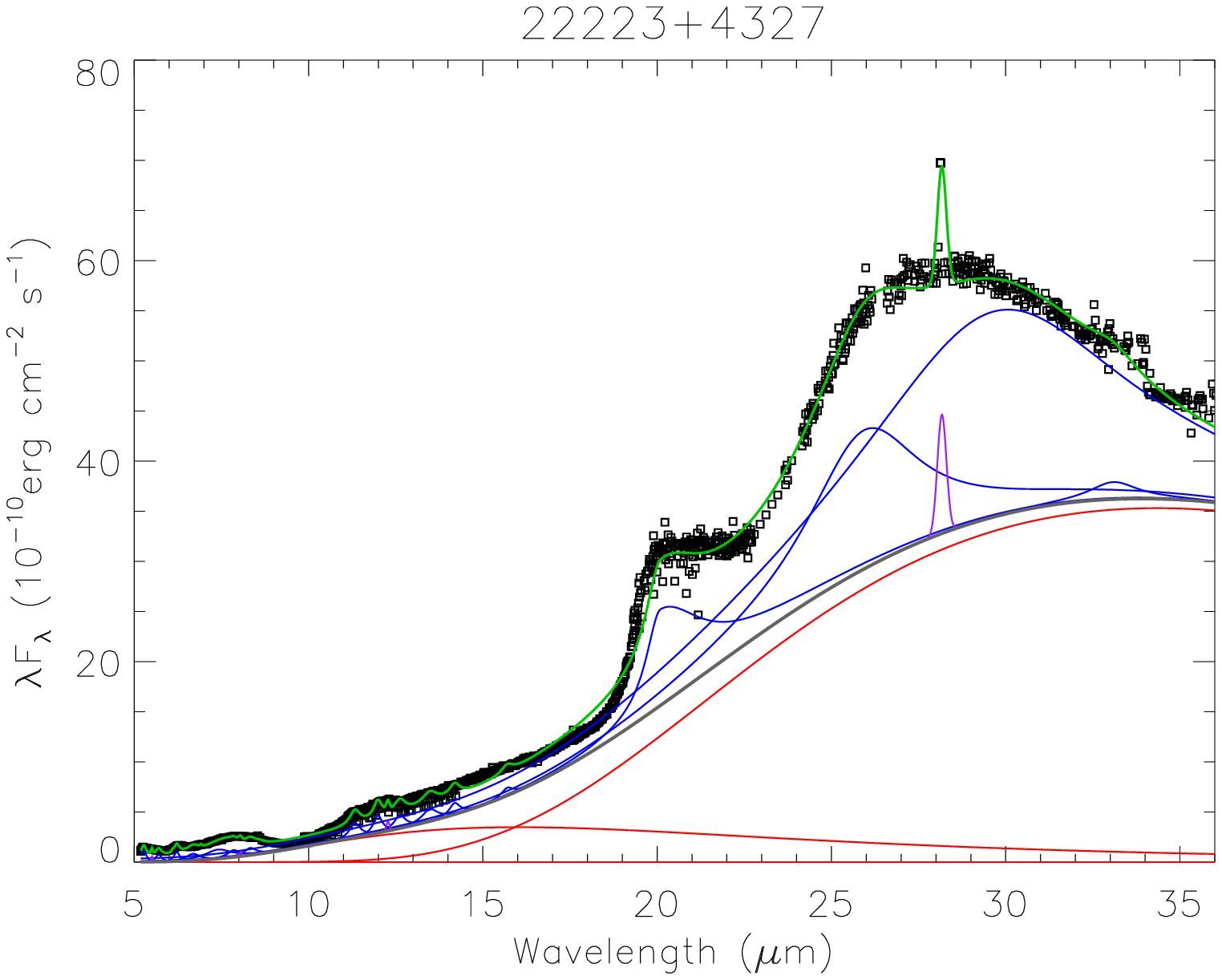}} \\
\resizebox{85mm}{!}{\includegraphics{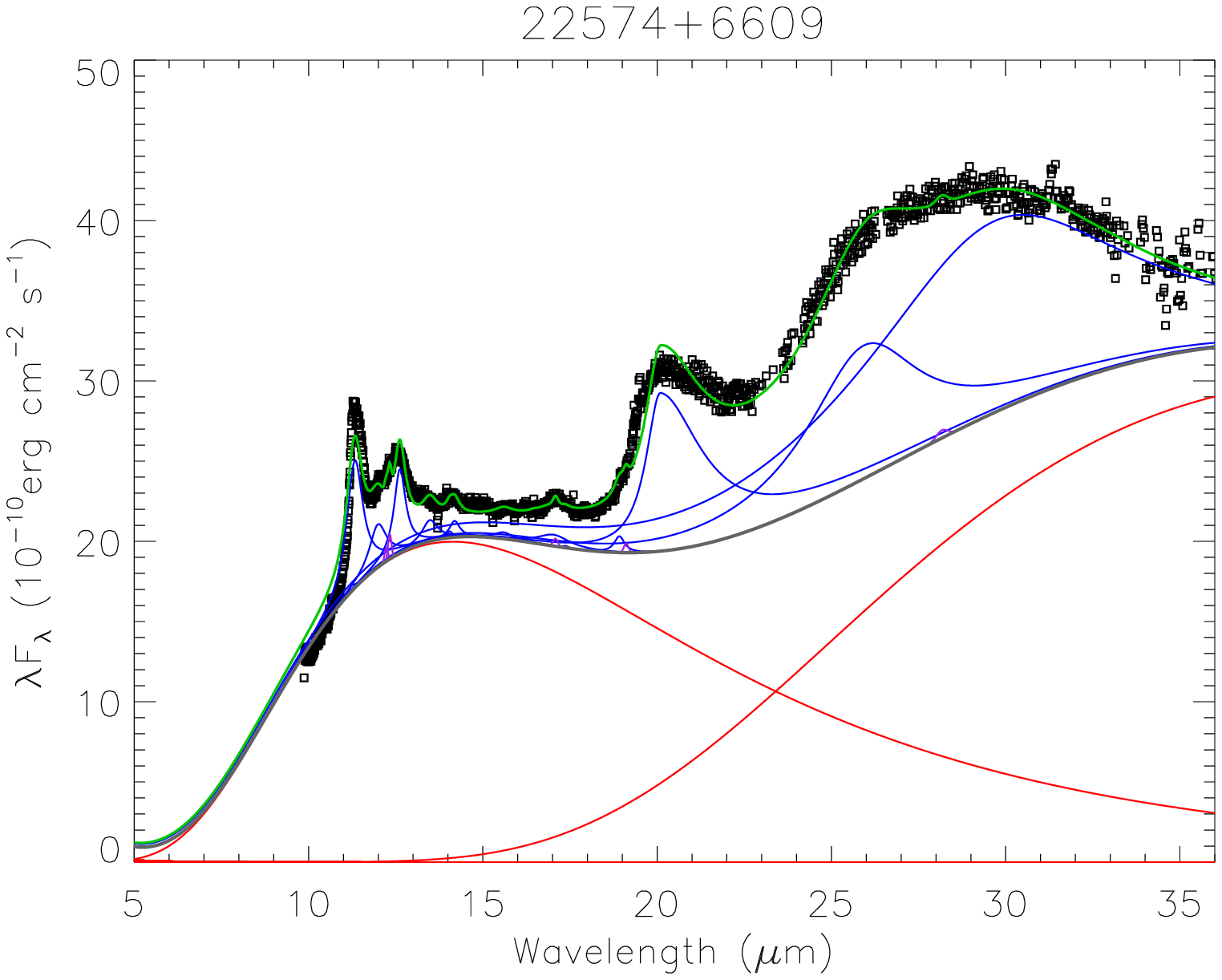}} & \resizebox{85mm}{!}{\includegraphics{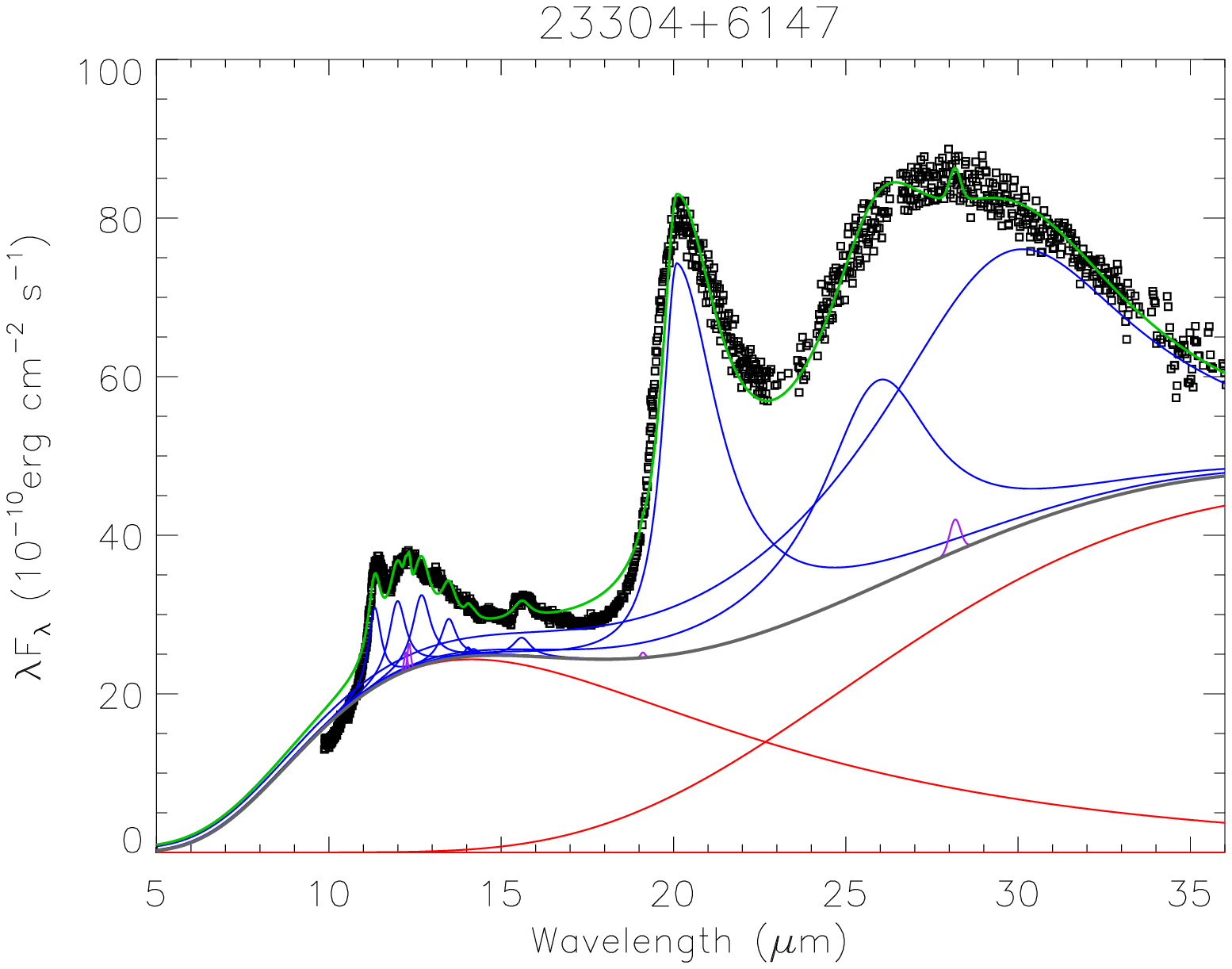}} \\
\end{tabular}
\end{center}
\caption{continued.}
\end{figure*}

\begin{figure*}
\epsfig{file=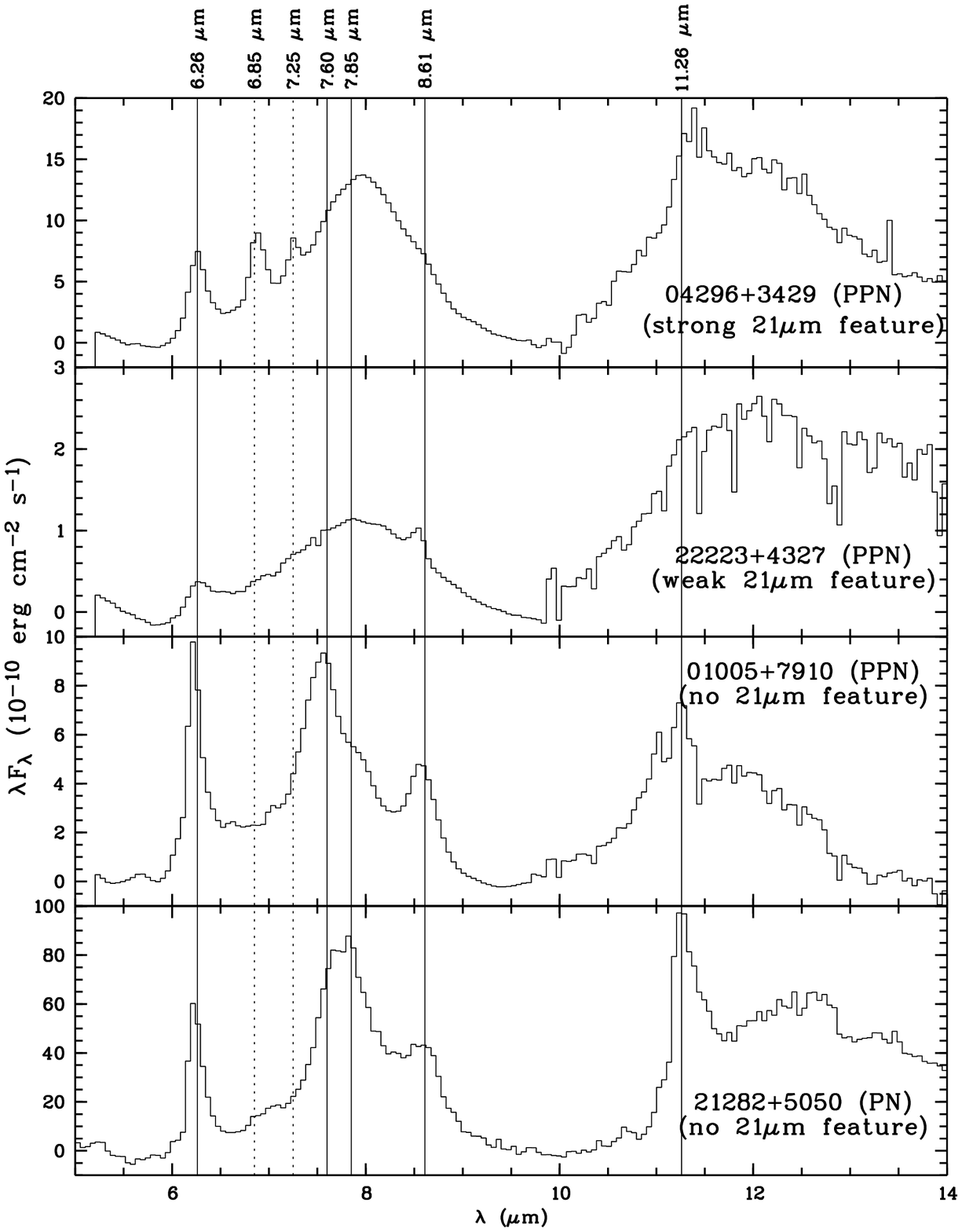,
height=20cm}
\caption{Continuum-subtracted IRS spectra. The continuum is the sum of the model stellar and dust continuum components.  The solid and dash lines indicate the positions of aromatic and aliphatic emission features, 
respectively.
}
\label{spe}
\end{figure*}

\begin{figure*}
\epsfig{file=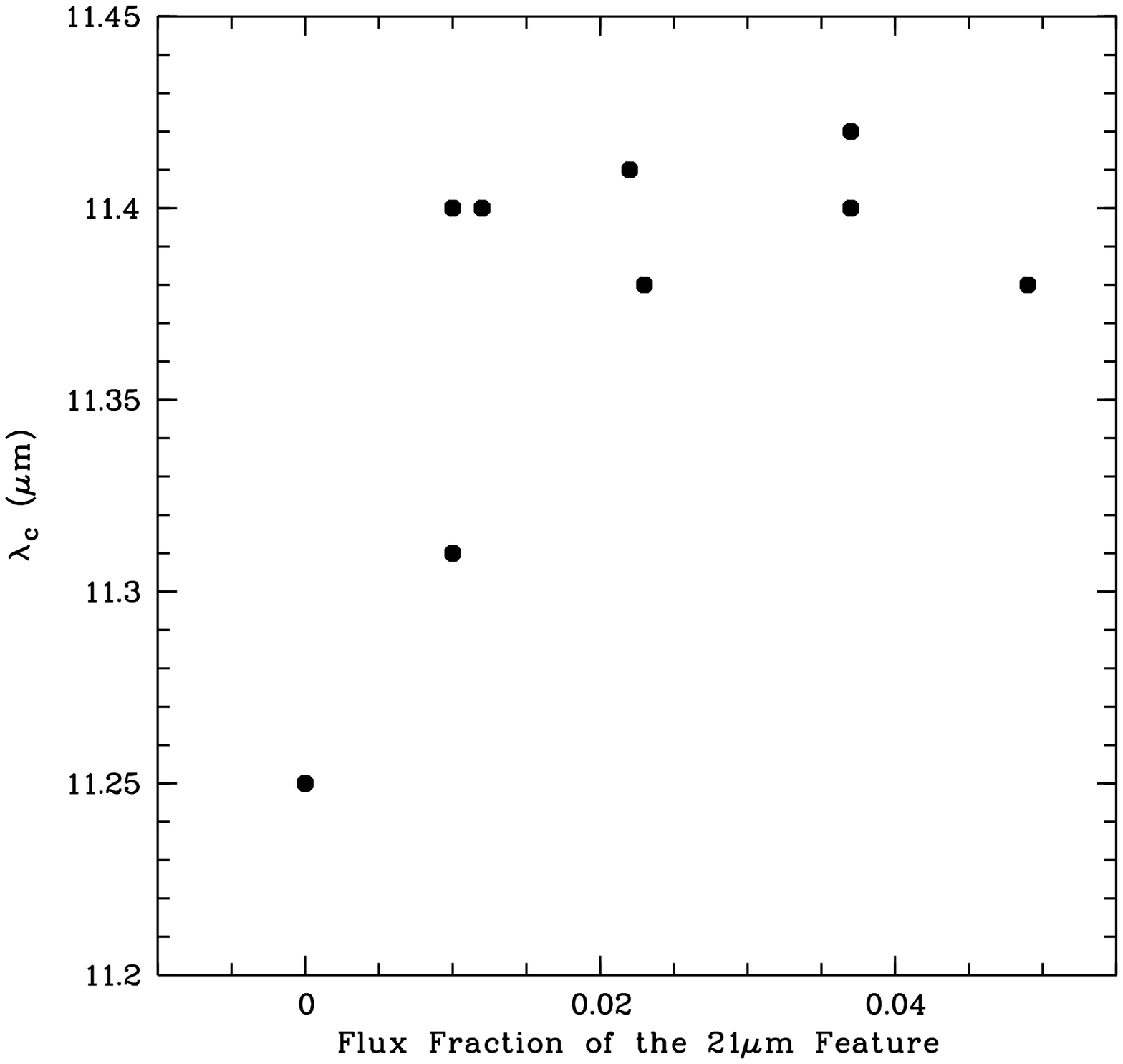,
height=12cm}
\caption{Peak wavelengths of the 11.3 $\mu$m feature versus flux fractions of the 21\,$\mu$m feature.  The object with zero 21 $\mu$m flux is IRAS 01005+7910.
}
\label{peak}
\end{figure*}

\begin{figure*}
\epsfig{file=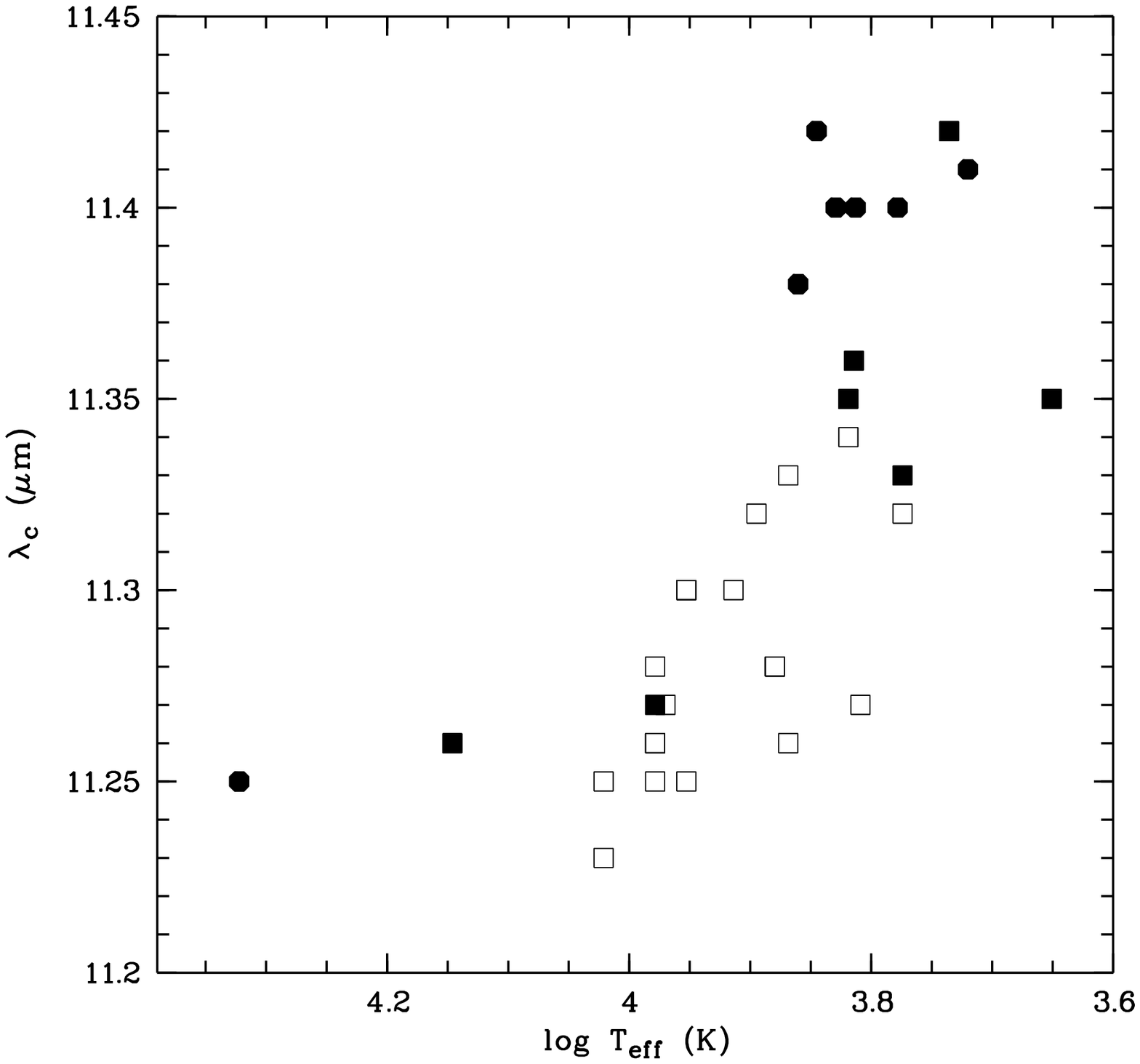,
height=12cm}
\caption{Peak wavelength of the 11.3 $\mu$m feature versus effective temperatures of the central stars.  The filled squares are objects from \citet{slo07}, the open squares are Ae/Be stars from \citet{keller08} and the filled circles are PPNs in the present study.
}
\label{peak2}
\end{figure*}

\begin{figure*}
\epsfig{file=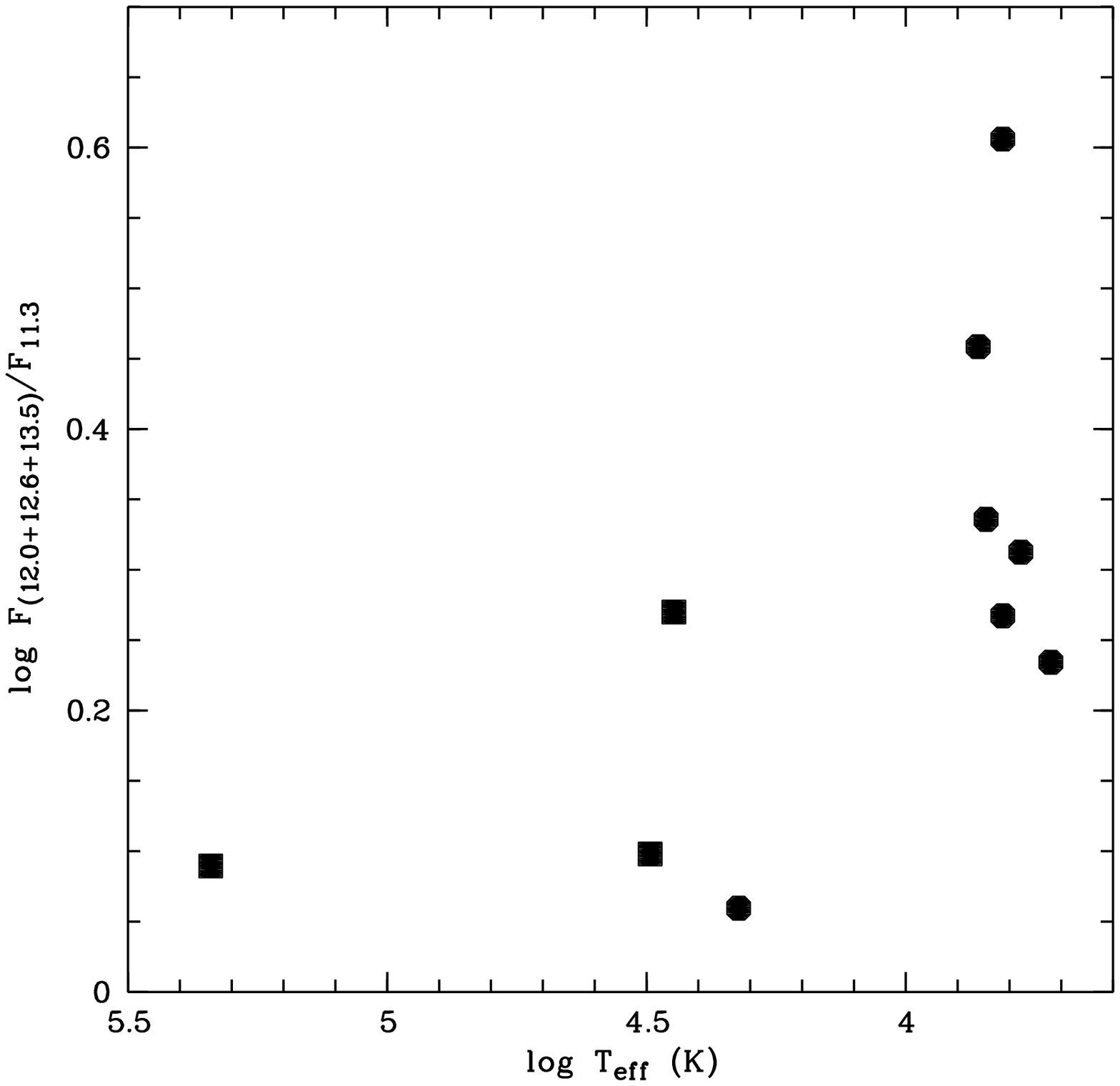,
height=12cm}
\caption{Ratio of the sum of the strengths of the 12.0, 12.6, and 13.5 $\mu$m features to the strength of the 11.3 $\mu$m feature as a function of central star temperature.  The filled circles are PPNs in our sample, and the filled squares are (from right to left) PNs IRAS 21282+5050, BD+303639, and NGC 7027.
}
\label{duo}
\end{figure*}

\begin{figure*}
\epsfig{file=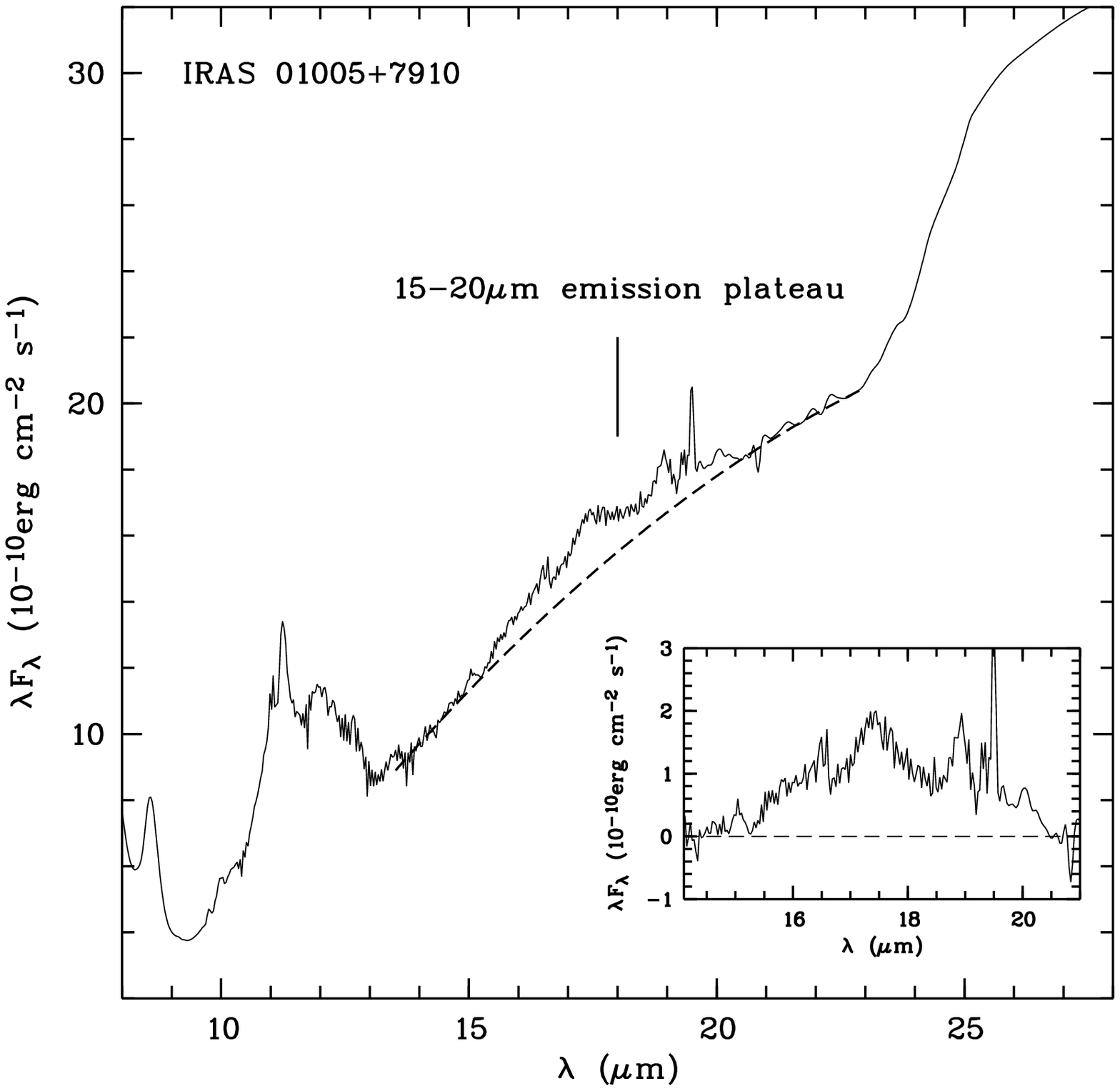, height=15cm}
\caption{The 15--20\,$\mu$m emission plateau. The dashed line
indicates the continuum.  The continuum-subtracted spectrum is
shown in the lower right panel. }
\label{plateau}
\end{figure*}

\begin{figure*}
\epsfig{file=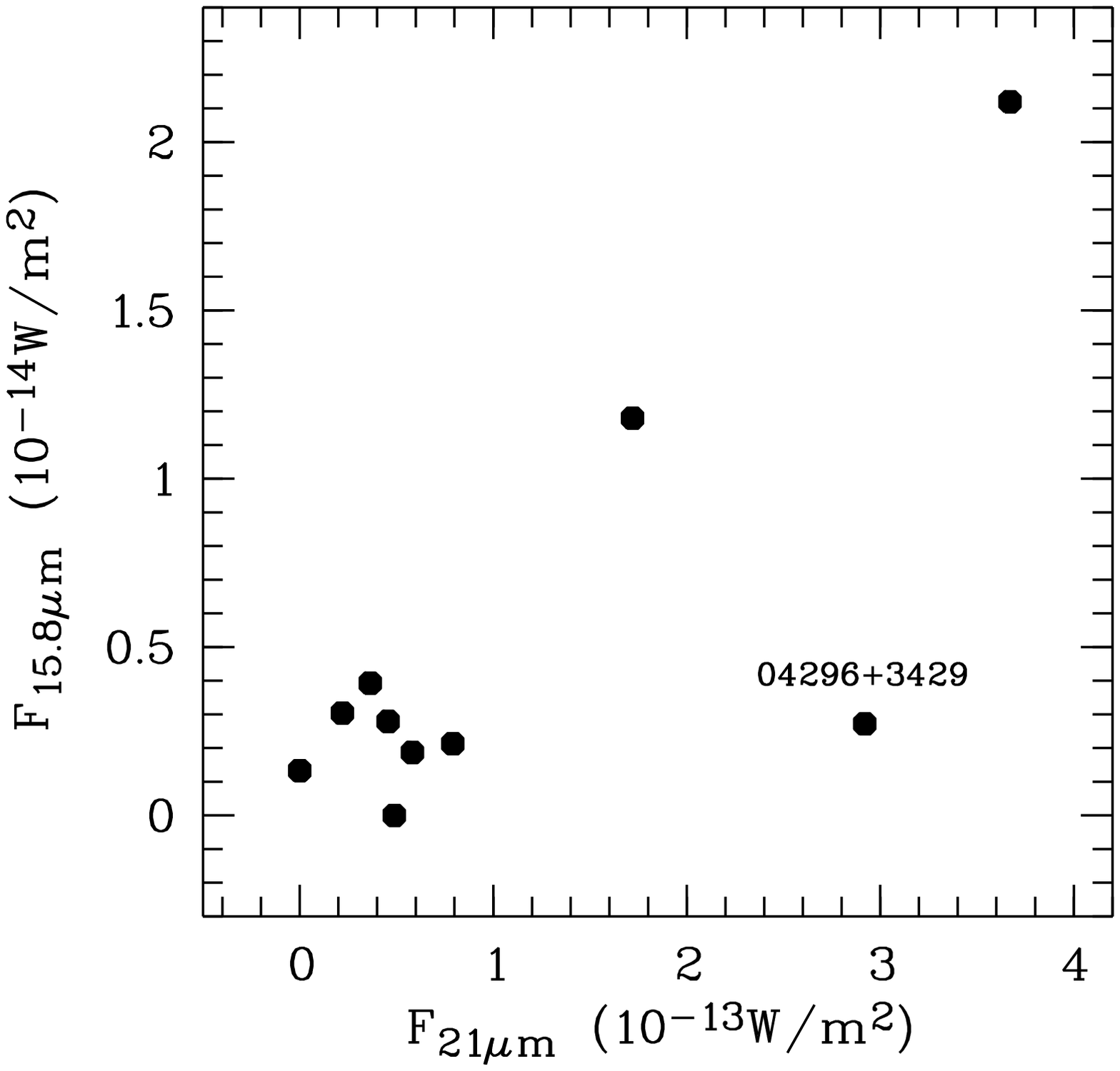,
height=12cm}
\caption{The strength of the 15.8 $\mu$m feature versus the strength of the
21 $\mu$m feature.
}
\label{vs1521}
\end{figure*}

\begin{figure*}
\epsfig{file=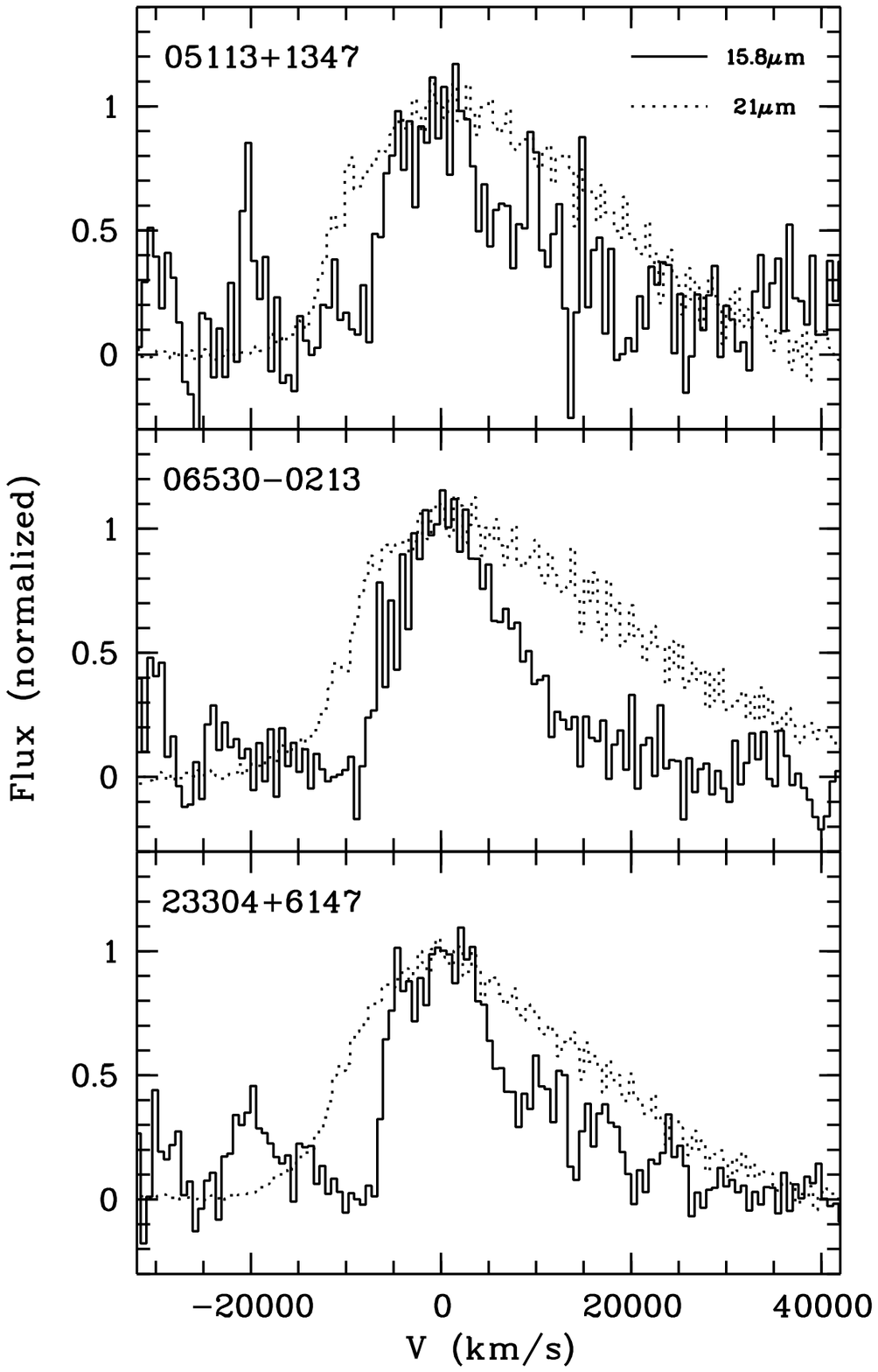, height=20cm}
\caption{The profiles of the 15.8 and 21\,$\mu$m features in three
PPNs.  }
\label{cor1521}
\end{figure*}

\begin{figure*}
\epsfig{file=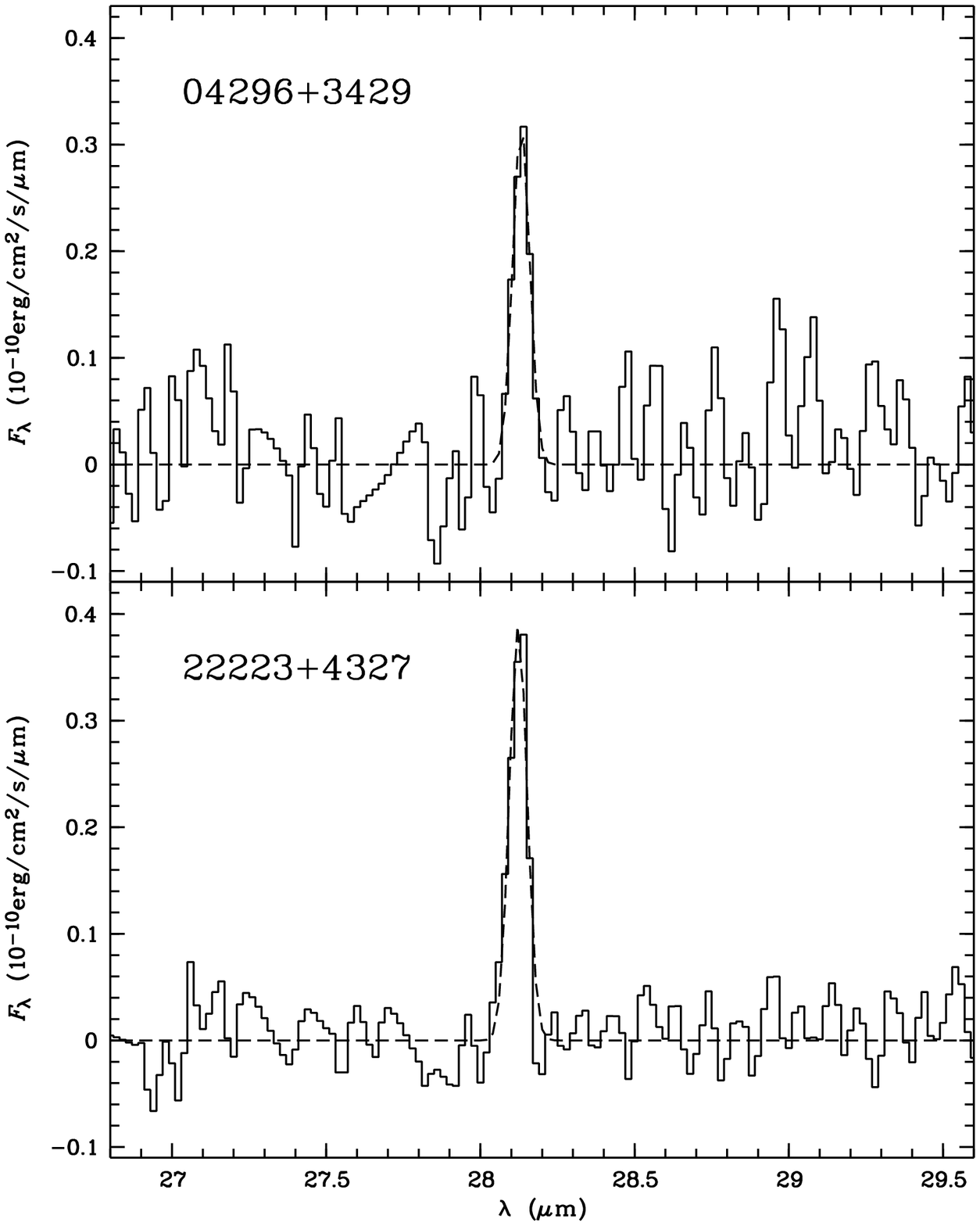, height=20cm}
\caption{The observed profiles of the H$_2$ 0--0 S(0) transitions in IRAS 04296+3429 and IRAS 22223+4327.
The dashed lines represent a Gaussian fit.}
\label{hyd}
\end{figure*}

\begin{figure*}
\epsfig{file=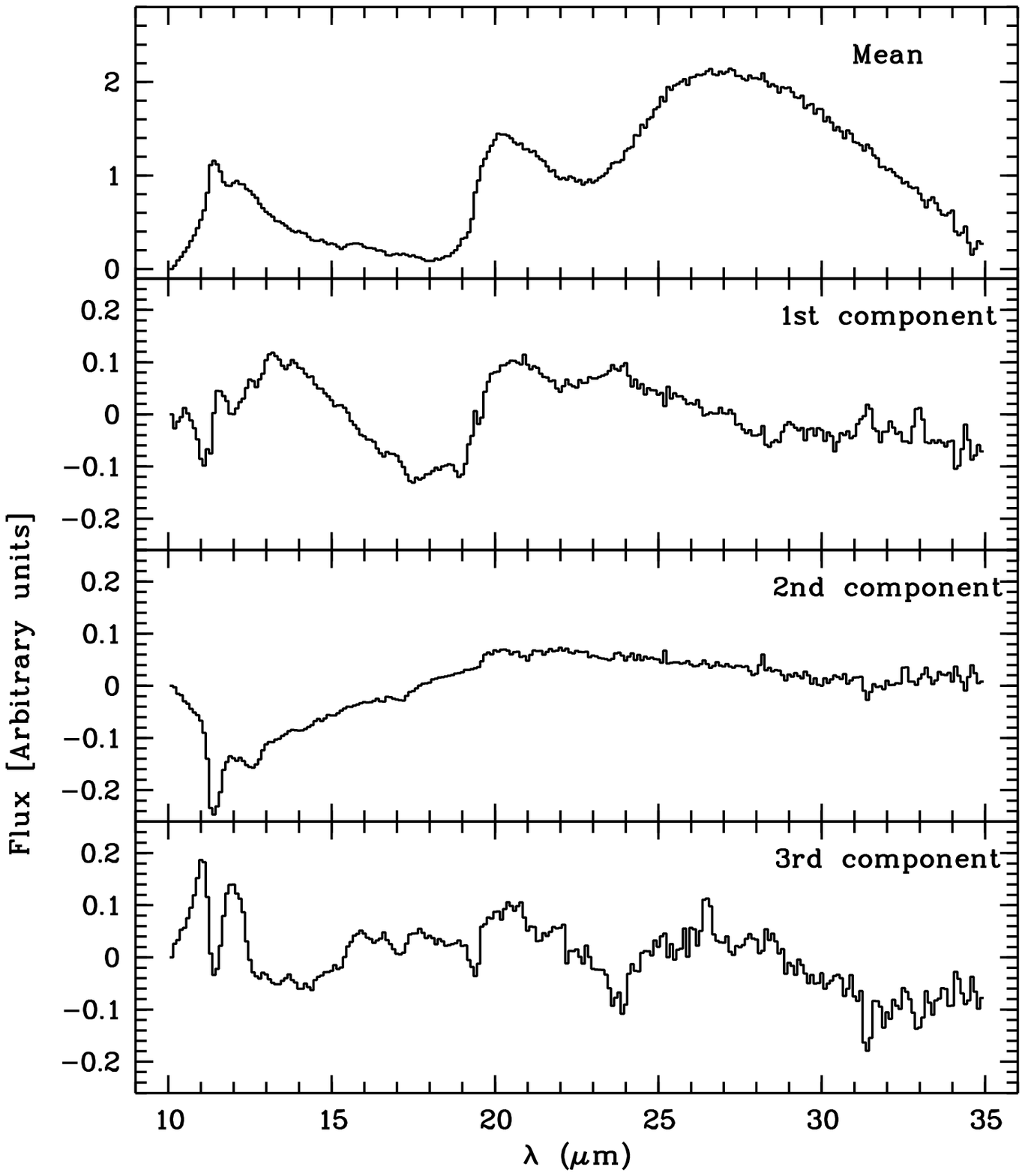, height=15cm}
\caption{The mean spectrum and the first three PCs.}
\label{pca}
\end{figure*}

\begin{figure*}
\epsfig{file=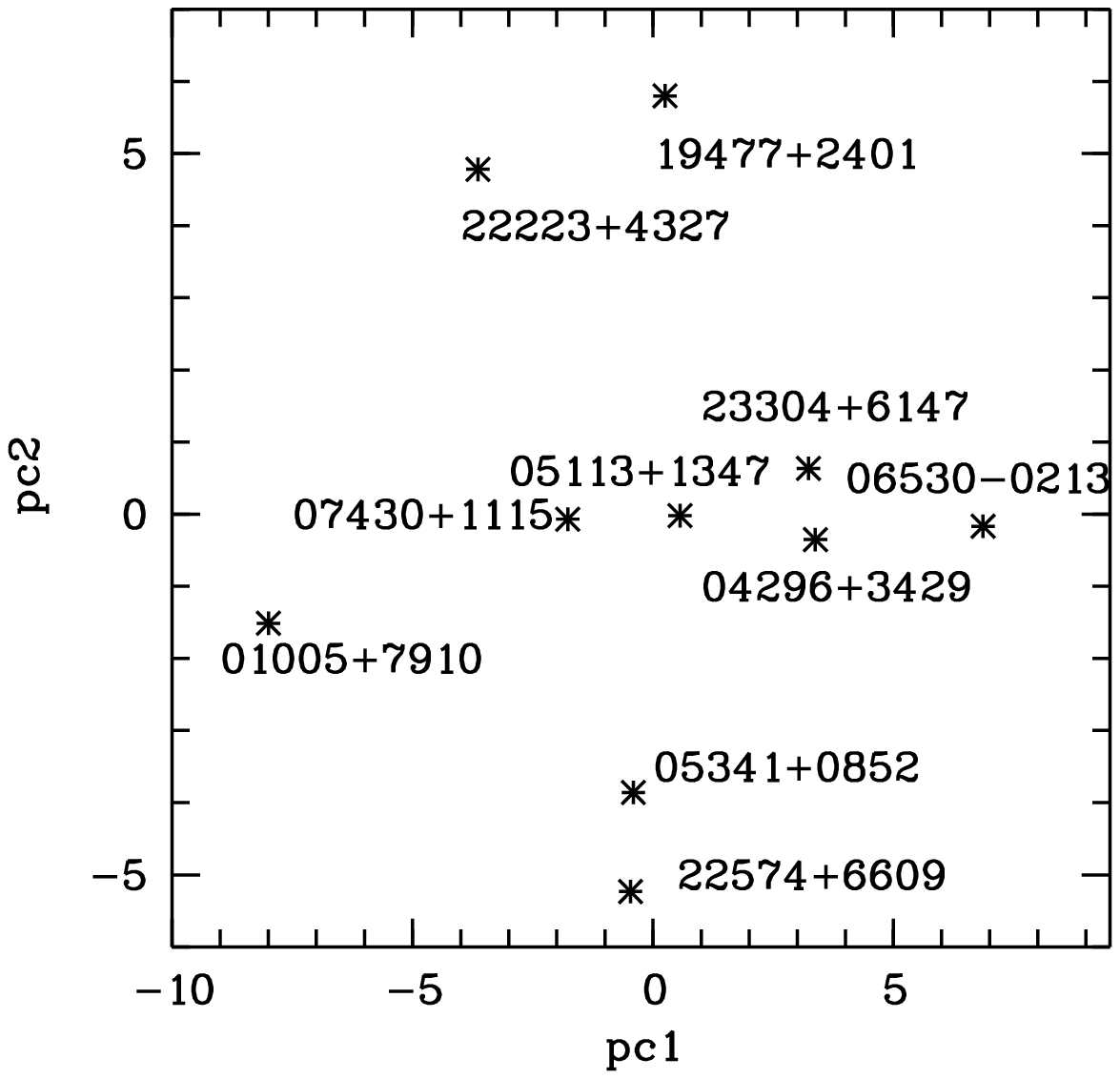, height=14cm}
\caption{Projections on to the first and the second eigenvectors of the spectrum sample.}
\label{pca_wei}
\end{figure*}

\clearpage
\newpage

%%%%%%%%%%%%
\begin{deluxetable}{cccc@{\extracolsep{0.1in}}cc@{\extracolsep{0.1in}}cc@{\extracolsep{0.1in}}cc@{\extracolsep{0.1in}}cc}
\rotate
\tablecaption{ Effective temperatures and integrated intensities of the dust continuum and
the 21 and 30\,$\mu$m features. \label{dust} }
\tabletypesize{\scriptsize}
\tablewidth{0pt}
\tablehead{ \colhead{Object}&
\colhead{Stellar}&
\multicolumn{4}{c}{Dust Continuum} &
\multicolumn{6}{c}{Dust Features$^b$} \\
\cline{3-6} \cline{7-12}
\colhead{IRAS ID} &
\colhead{Temperature$^a$} &
\multicolumn{2}{c}{Warm} &
\multicolumn{2}{c}{Cold}&
\multicolumn{2}{c}{20.1\,$\mu$m} &
\multicolumn{2}{c}{26\,$\mu$m} &
\multicolumn{2}{c}{30\,$\mu$m} \\
\cline{3-4} \cline{5-6} \cline{7-8} \cline{9-10} \cline{11-12}
\colhead{} &
\colhead{(K)} &
\colhead{(K)} &
\colhead{(W/m$^2$)} &
\colhead{(K)} &
\colhead{(W/m$^2$)} &
\colhead{(W/m$^2$)} &
\colhead{($\%$)$^c$} &
\colhead{(W/m$^2$)} &
\colhead{($\%$)$^c$} &
\colhead{(W/m$^2$)} &
\colhead{($\%$)$^c$} \\ }
\startdata
$01005+7910$ & 21000 &140&1.12E$-12$ & 70&1.38E$-12$ & 0.00(0.00)E$-00$&0.0 & 1.30(0.15)E$-13$& 4.0 & 5.47(0.21)E$-13$&16.7\\
$04296+3429$ &  7000 &180&2.39E$-12$ & 60&3.15E$-12$ & 2.92(0.01)E$-13$&3.7 & 6.19(0.04)E$-13$& 7.9 & 1.12(0.01)E$-12$&14.3\\
$05113+1347$ &  5250 &150&6.49E$-13$ & 70&8.90E$-13$ & 4.57(0.06)E$-14$&2.2 & 1.57(0.04)E$-13$& 7.5 & 4.12(0.11)E$-13$&19.7\\
$05341+0852$ &  6500 &180&6.46E$-13$ & 70&7.52E$-13$ & 2.20(0.06)E$-14$&1.2 & 9.25(0.35)E$-14$& 5.1 & 1.88(0.08)E$-13$&10.3\\
$06530-0213$ &  7250 &180&9.50E$-13$ & 70&1.80E$-12$ & 1.72(0.01)E$-13$&4.9 & 1.80(0.03)E$-13$& 5.1 & 2.43(0.08)E$-13$& 6.9\\
$07430+1115$ &  6000 &180&3.13E$-13$ & 70&2.22E$-12$ & 3.64(0.06)E$-14$&1.0 & 4.09(0.04)E$-13$&10.8 & 6.53(0.08)E$-13$&17.2\\
$19477+2401$ &...$^d$&...&...        & 80&1.62E$-12$ & 4.89(0.06)E$-14$&2.3 & 1.82(0.04)E$-13$& 8.7 & 2.19(0.08)E$-13$&10.5\\
$22223+4327$ &  6500 &150&4.03E$-13$ & 70&3.89E$-12$ & 7.91(0.09)E$-14$&1.4 & 3.66(0.04)E$-13$& 6.6 & 7.30(0.07)E$-13$&13.2\\
$22574+6609$ &...$^e$&170&2.07E$-12$ & 60&3.25E$-12$ & 5.82(0.01)E$-14$&1.0 & 1.53(0.04)E$-13$& 2.5 & 3.30(0.08)E$-13$& 5.5\\
$23304+6147$ &  6750 &170&2.56E$-12$ & 60&5.07E$-12$ & 3.67(0.01)E$-13$&3.7 & 5.78(0.04)E$-13$& 5.8 & 1.10(0.01)E$-12$&11.1\\
\enddata
\tablenotetext{a}{Based on published model atmosphere studies.}
\tablenotetext{b}{The numbers in brackets are fitting uncertainties.}
\tablenotetext{c}{The flux fractions of the features to the total IR emission, which includes
the contribution from the dust features from 10--35\,$\mu$m and the dust
thermal continuum.}
\tablenotetext{d}{G0 spectral type.}
\tablenotetext{e}{A1 spectral type. }
\end{deluxetable}
%%%%%%%%%%%%%%%%%%%%%%

\textheight 24.5cm

\begin{deluxetable}{cccccccccccccc}
\rotate
\tablecaption{Strengths of AIB and 15.8 $\mu$m features.$^a$ \label{AIB}}
\tabletypesize{\scriptsize}
\tablewidth{9.8in}
\tablehead{
\colhead{ IRAS ID} &
\colhead{6.2\,$\mu$m} &
\colhead{6.9\,$\mu$m} &
\colhead{7.4\,$\mu$m} &
\colhead{7.8\,$\mu$m} &
\colhead{8.3\,$\mu$m} &
\colhead{8.6\,$\mu$m} &
\colhead{10.7\,$\mu$m} &
\colhead{11.3\,$\mu$m} &
\colhead{12.0\,$\mu$m} &
\colhead{12.6\,$\mu$m} &
\colhead{13.5\,$\mu$m} &
\colhead{15.8\,$\mu$m} &
\colhead{\% of total$^c$}\\
}
\startdata
$01005+7910$&3.84(0.12)&0.73(0.02)&9.58(0.40)&1.58(0.23)&1.32(0.18)&2.72E(0.14)&1.03(0.20)&2.75(0.38)&2.92(0.33)&0.23(0.06)&0.00(0.00)&0.13(0.01)&8.2\\
$04296+3429$&3.71(0.12)&4.82(0.19)&17.70(3.00)&8.70(0.19)&6.68(0.18)&2.29E(0.14)&1.09(0.23)&8.26(0.27)&8.54(0.34)&6.21(0.31)&3.15(0.25)&0.27(0.02)&9.1\\
$05113+1347$& ...$^b$        &...             &...             &...             &...             &...             &0.18(0.02)&2.25(0.28)&3.28(0.33)&0.58(0.02)&0.00(0.00)&0.28(0.02)&$>3.1$\\
$05341+0852$& ...      &...       &...       &...       &...       &...       &0.23(0.02)&3.62(0.28)&3.00(0.34)&2.23(0.33)&1.47(0.27)&0.30(0.02)&$>5.9$\\
$06530-0213$& ...      &...       &...       &...       &...       &...       &0.00(0.00)&3.57(0.26)&4.13(0.34)&3.84(0.31)&2.29(0.26)&1.18(0.16)&$>4.3$\\
$07430+1115$& ...      &...       &...       &...       &...       &...       &0.99(0.02)&4.83(0.28)&5.79(0.34)&2.71(0.33)&1.42(0.26)&0.39(0.02)&$>4.2$\\
$19477+2401^d$& ...      &...       &...       &...       &...       &...       &0.15(0.03)&0.57(0.04)&0.53(0.04)&0.69(0.03)&0.26(0.03)&0.00(0.00)&$>1.0$\\
$22223+4327$&0.29(0.01)&0.54(0.02)&1.79(0.42)&0.63(0.02)&0.55(0.02)&0.17(0.01)&0.00(0.00)&0.63(0.04)&1.08(0.33)&0.70(0.03)&0.75(0.03)&0.21(0.01)&1.3\\
$22574+6609$& ...      &...       &...       &...       &...       &...       &0.00(0.00)&5.55(0.26)&2.71(0.34)&3.68(0.33)&1.10(0.27)&0.19(0.02)&$>2.2$\\
$23304+6147$& ...      &...       &...       &...       &...       &...       &0.00(0.00)&2.64(0.15)&9.49(0.32)&6.87(0.32)&4.31(0.26)&2.12(0.14)&$>2.6$\\
\enddata
\tablenotetext{a}{ In unit of 10$^{-14}$W/m$^2$. The numbers in brackets are fitting uncertainties.}
\tablenotetext{b}{no feature strengths derived for the 7 objects without SL observations.}
\tablenotetext{c}{The flux fractions are the ratios of the sum of all the AIB features to the total IR emission, which includes the contribution from the dust features from 10--35\,$\mu$m and the dust
thermal continuum.}
\tablenotetext{d}{The flux measurements of this object may not be reliable (see section 2).}
\end{deluxetable}

\clearpage
\begin{deluxetable}{ccccc}
\tablecaption{Peak wavelengths of AIB features.$^a$ \label{wave} }
\tabletypesize{\scriptsize}
\tablewidth{0pt}
\tablehead{
\colhead{IRAS ID} &
\colhead{6.2\,$\mu$m} &
\colhead{7.7--8.2\,$\mu$m} &
\colhead{8.6\,$\mu$m} &
\colhead{11.3\,$\mu$m} \\
}
\startdata
$01005+7910$ &  $6.20\pm0.03$ &  $7.57\pm0.05$  &  $8.56\pm0.04$ &  $11.25\pm0.02$ \\
$04296+3429$ &  $6.26\pm0.02$ &  $8.02\pm0.08$  &  ...           &  $11.42\pm0.09$ \\
$05113+1347$ &   ...       &  ...         &  ...                 &  $11.41\pm0.13$ \\
$05341+0852$ &   ...       &  ...         &  ...                 &  $11.40\pm0.09$ \\
$06530-0213$&   ...       &  ...         &  ...                 &  $11.38\pm0.02$ \\
$07430+1115$ &   ...       &  ...         &  ...                 &  $11.40\pm0.13$ \\
$19477+2401$ &   ...       &  ...         &  ...                 &  $11.38\pm0.13$ \\
$22223+4327$ &  $6.29\pm0.03$ &  $7.87\pm0.04$  &  $8.53\pm0.03$ &  $>11.30$     \\
$22574+6609$ &   ...       &  ...         &  ...                 &  $11.31\pm0.04$ \\
$23304+6147$ &   ...       &  ..          &  ...                 &  $11.40\pm0.06$\\
\enddata   
\tablenotetext{a}{ In unit of $\mu$m.}
\end{deluxetable}

\begin{deluxetable}{cccc}
\tablecaption{ H$_2$ lines in IRAS~04296+3429 and IRAS~22223+4327. \label{h2} }
\tabletypesize{\scriptsize}
\tablewidth{0pt}
\tablehead{
\colhead{IRAS ID} &
\colhead{$F$(0--0 S(0))} &
\colhead{$F$(0--0 S(1))} &
\colhead{$T_{\rm ex}$}\\
&
\colhead{(W/m$^2$)} &
\colhead{(W/m$^2$)} 
&\colhead{(K)}\\
}
\startdata
$04296+3429$ & 3.2E$-14$ & $<4$E$-16$ & $<56$ \\
$22223+4327$ & 3.9E$-14$ & $<2$E$-16$ & $<51$ \\
\enddata
\end{deluxetable}

\end{document}